\documentclass[fleqn,usenatbib]{mnras}
\usepackage[english]{babel}

\usepackage[T1]{fontenc}
\usepackage{ae,aecompl}

\usepackage{fancyhdr}
\usepackage{amsfonts}
\usepackage{amsmath}
\usepackage{amssymb}
\usepackage{multicol}
\usepackage{layout}
\usepackage{graphicx}
\usepackage{multirow}
\usepackage{color}
\usepackage{multirow}
\usepackage{times}
\usepackage{natbib}
\def\be{\begin{equation}}
\def\ee{\end{equation}}
\newcommand{\jt}[1]{{\textcolor{black}{#1}}}

\newif\ifAMStwofonts
\AMStwofontstrue

\graphicspath{{./fig/}}

\newcommand{\CC}{\Lambda}

\title[Cosmography to RVMs]{Cosmographic approach to Running Vacuum  dark energy models: new constraints using BAOs and Hubble diagrams at higher redshifts}

\author[M. Rezaei et~al.]{Mehdi Rezaei$^{1}$,
	 {Joan Sol\`a Peracaula$^2$}, {Mohammad Malekjani$^{1}$ \thanks{malekjani@basu.ac.ir}}
	\\
	$^1$ Department of Physics, Bu Ali Sina University, Hamedan 65178, Iran.\\
	$^2$ Departament de F\'\i sica Qu\`antica i Astrof\'\i sica  and   Institute of Cosmos Sciences,  Universitat de Barcelona, Avinguda Diagonal 647 E-08028 Barcelona, Catalonia, Spain.\\
}

\date{Accepted ?, Received ?; in original form \today}

\pagerange{\pageref{firstpage}--\pageref{lastpage}} \pubyear{2021}

\begin{document}

\label{firstpage}
\maketitle

\begin{abstract}
In this work we study different types of dark energy (DE) models  in the framework of the cosmographic approach, with emphasis on the Running Vacuum models (RVMs). We assess their viability using different information criteria and compare them with the so-called Ghost DE models (GDEs) as well as with the concordance $\Lambda$CDM model.  We use the  Hubble diagrams for Pantheon SnIa, quasars (QSOs), gamma-ray bursts (GRBs) as well as the data on baryonic acoustic oscillations (BAOs) in four different combinations. Upon minimizing the $\chi^2$ function of the distance modulus in the context of the Markov Chain Monte Carlo method (MCMC), we put constraints on the current values of the standard cosmographic parameters in a model-independent way.   It turns out that, in the absence of BAOs data,  the various DE models generally exhibit cosmographic tensions with the observations at the highest redshifts (namely with the QSOs and GRBs data).  However,  if we include the robust observations from BAOs to our  cosmographic sample,  the $\Lambda$CDM  and RVMs  are clearly favored against  the GDEs.  Finally, judging from the perspective  of the deviance information criterion (DIC), which enables us to compare models making use of the Markov chains of the  MCMC method,  we conclude that the RVMs  are the preferred kind of DE models. We find it remarkable that these models,  which  had been previously shown to be capable of alleviating  the $\sigma_8$ and $H_0$ tensions,  appear now also as the most successful ones at the level of the cosmographic analysis.
	
\end{abstract}
\begin{keywords}
	Cosmology: dark energy.
\end{keywords}
\maketitle

\section{Introduction}

From a theoretical point of view, the accelerated expansion of our Universe can be caused by a simple cosmological constant term or, more generally,  by an exotic component $X$ with negative pressure which violates the strong energy condition, $\rho_X +3 p_X > 0$,  $\rho_X + p_X>0$,  while still preserving the weak one: $\rho_X+p_X\geq0$,  $\rho_X\geq0$. Such a generic component would have an equation of state $p_X=w_X\rho_X$ with $w_X<-1/3$  and $\rho_X\geq0$, thus producing the necessary speeding up of the Universe's expansion\,\citep{Turner:1998ex}.  Many other possibilities are of course available, some of them to be discussed below. Extensive evidence of such cosmic acceleration has been first presented through the luminosity distances of type Ia supernovae\,\citep{Riess1998,Perlmutter1999,Kowalski2008,Scolnic:2017caz} and of course in modern times through the precise measurements of the  anisotropies of the cosmic microwave background (CMB)\,\citep{Aghanim:2018eyx}. Due to the fact that most properties of such an exotic component remain unknown, it has been dubbed the dark energy (DE). The results of different cosmological observations indicate that the current universe is spatially flat and dark energy occupies approximately $70 \%$ of the energy budget of the universe \citep{Spergel:2003cb,Peiris:2003ff,Bennett:2003bz,Aghanim:2018eyx}. The first theoretical candidate for dark energy is  the cosmological constant, $\Lambda$, with the equation of state (EoS) parameter equal to $-1$.  The $\CC$-cosmology (also called the standard or concordance cosmological model) is the framework  assuming the presence of a cosmological constant $\Lambda$ and cold dark matter (CDM), and  for this reason is also denoted as $\Lambda$CDM.  It is  mostly consistent with  the current cosmological observations, but not quite.  In addition, from the theoretical perspective it also suffers from two serious conundrums: the fine-tuning and the cosmic coincidence problems\,\citep{Weinberg1989,Padmanabhan2003,Copeland2006}. However, these conundrums are not restricted to $\CC$, they ultimately affect all known forms of DE in one way or another\,\citep{Sola:2013gha}.

Here we concentrate on the practical problems of the $\CC$CDM related with observations.  In the last few years a bunch of  observational data indicate that the $\CC$-cosmology is afflicted with some worrisome discrepancies concerning the prediction of relevant cosmological parameters. In particular,  the current value of the Hubble constant $H_0$ as measured locally and the one extracted from the sound horizon obtained from  the cosmic microwave background (CMB) are in significant tension\,\citep{Riess:2019cxk}. These values provide the two chief absolute scales at opposite ends of the visible expansion history of the Universe\,\citep{DiValentino:2020zio}. Comparing these two scales gives a stringent end-to-end test of $\Lambda$ cosmology, over the full history of the Universe \citep{Verde:2019ivm}. By improving the accuracy of the $H_0$ measurement from Cepheid-calibrated type Ia supernovae (SnIa), evidence has been growing of a significant discrepancy between the two. The local and direct determination gives $H_0=74.03\pm 1.42$ km/s/Mpc from \citep{Riess:2019cxk} while the one inferred from Planck CMB data using $\Lambda$CDM cosmology  predicts $H_0 = 67.4\pm0.5$ km/s/Mpc \citep{Aghanim:2018eyx},  whence a significant discrepancy of  $4.4 \sigma$.  Such tension may well be passing the point of being attributable to a fluke, as emphasized by\,\citep{Riess:2019cxk}. The other significant (though lesser) tension concerns the discrepancy between the amplitude of matter fluctuations from large scale structure (LSS) data \citep{Macaulay:2013swa}, and the value predicted by CMB experiments based on the $\Lambda$CDM, see\,\citep{DiValentino:2020vvd} for an updated status. Let us also mention that  the Baryon Acoustic Oscillations (BAOs) measured by the Lyman-$\alpha$ forest, reported in \citep{Delubac:2014aqe}, predict a smaller value of the matter density parameter ($\Omega_{\rm m}$) compared with those obtained from CMB data. For a recent and detailed discussion about the current challenges for the $\CC$CDM, see \cite{Perivolaropoulos:2021jda}, and references therein.

To overcome the problems of $\Lambda$-cosmology (i.e. the $\CC$CDM model) or at least  to alleviate some the above tensions, different kinds of DE frameworks  have been investigated in the literature \citep{Caldwell:1997ii,Erickson:2001bq,Armendariz2001,Caldwell2002,Padmanabhan2002,Elizalde:2004mq}.  For a summary of the many possible models competing to solve the mentioned tensions, see e.g. \citep{DiValentino:2019jae,DiValentino:2021izs,Perivolaropoulos:2021jda} and references therein. Some of these models have good consistency with different observations while some others have been ruled out after comparing with observational data \citep[see also][]{Malekjani:2016edh,Rezaei:2017yyj, Rezaei:2017hon,Malekjani:2018qcz,Lusso:2019akb,Rezaei:2019roe,Rezaei:2019hvb,Lin:2019htv,Rezaei:2019xwo,Rezaei:2020mrj,Rezaei:2021qpq}.
Therefore, cosmologists felt motivated to propose different scenarios in which the origin of DE is based on more physical principles and ultimately on fundamental theory. For instance, in \citep{Shapiro:2000dz} it was  proposed a possible connection between cosmology and quantum field theory (QFT) on the basis of the renormalization group, see also \citep{Babic:2004ev}.  This gave rise to the idea of running vacuum models (RVMs), cf. \,\cite{Sola:2013gha} and references therein for an ample exposition.  The effective form of the vacuum energy density for the RVM was later generalized as follows\,\citep{Sola:2015rra}:
\begin{equation}\label{rLRVM}
\rho_{\rm RVM}(H) = \frac{\Lambda(H)}{\kappa^2}=
\frac{3}{\kappa^2}\left(c_0 + \nu H^{2} +\tilde{\nu} \dot{H} +
\frac{H^{4}}{H_{I}^{2}} + \dots \right) \,,
\end{equation}
in which  \dots stand for higher order (even) powers of $H$ beyond four;   $\kappa^2=8\pi G$,  and $c_0 >0$ is a constant closely related to the (current era) cosmological constant. In the above expression, $H_I$ denotes the Hubble scale at the inflationary epoch. The coefficients $\nu$, $\tilde{\nu}$  are small since they control the (mild) dynamics of the vacuum energy density.  When these coefficients vanish, the above expression boils down to the conventional  constant value of the  vacuum energy density in the $\CC$CDM: $\rho_{\Lambda}=\frac{3}{\kappa^2}\,c_0$, in which the cosmological constant is precisely $\Lambda=3c_0$.  However,  the RVM density receives in general a mild dynamical contribution of ${\cal O}(H^2)$, plus higher order terms which are irrelevant for the current universe. Thereby the effective cosmological term  $\Lambda(H)=3 c_0+{\cal O}(H^2)+...$   is no longer constant in the RVM.
The  corresponding equation of state is that of an ideal de-Sitter fluid, despite the time dependence of the vacuum energy:
\begin{align}\label{rvmeos}
p_{\rm RVM} (H) = - \rho_{\rm RVM} (H)~,
\end{align}
where $p_{\rm RVM} (H)$ denotes the vacuum pressure density.  As indicated before, around the current epoch and for all the practical cosmographic considerations relevant to the present study, it is sufficient to approximate the expression \eqref{rLRVM} by ignoring all the terms ${\cal O}(H^4)$. The latter, however, are crucial to account for inflation in the early universe within the context of the RVM\,\citep{Lima:2012mu,Perico:2013mna,Sola:2015rra,Sola:2015csa,SolaPeracaula:2019kfm}. Coefficients $\nu$ and  $\tilde{\nu}$  will be sensitive to our fits, as we shall see in the course of our analysis. It is remarkable that the RVM structure \eqref{rLRVM} is very well motivated theoretically. Apart from the mentioned old qualitative renormalization group arguments, the $\sim H^2$ terms have recently been  derived  from explicit computations of quantum effects of the effective action of  QFT in curved spacetime \citep{Moreno-Pulido:2020anb}, and the $\sim H^4$ ones can be accounted from string theory considerations, see\,\citep{Mavromatos:2020kzj} for a review.

In general, dynamical dark energy models (DDEs) can be helpful to improve the  fit the cosmological data, and in particular to alleviate the $H_0$ and $\sigma_8$ tensions.  Not all of  the DDEs, however,  are equally efficient in this task, see e.g.\,\citep{Sola:2018sjf,Gomez-Valent:2020mqn}. Recently,  torsional gravity has been invoked as a possible alleviation, in particular through the $f(T)$-class\,\citep{Anagnostopoulos:2019miu,Yan:2019gbw} although there is a large arbitrariness in the selection of $f(T)$. On the other hand,  the above mentioned `running vacuum models' (RVMs) \citep{Sola:2013gha}  seems to work optimally to mitigate these tensions,  as shown in our previous study \citep{Rezaei:2019xwo}. See also\,\citep{SolaPeracaula:2018xsi,Sola:2017znb} and in particular the latest fits testing the RVMs against the cosmological data as recently presented in \citep{Sola:2021txs}.  Previous devoted  analyses of the RVMs and their confrontation to the data can be found e.g. in \citep{Gomez-Valent:2014rxa,Gomez-Valent:2014fda,Sola:2015wwa,Sola:2016jky,Sola:2017jbl,Gomez-Valent:2015pia,Rezaei:2019xwo}.  The special theoretical status of the RVMs stems from the fact that their successful phenomenological performance is backed up by the aforementioned theoretical calculations  from QFT and string theory, which fix  their structure in the specific form Eq. \eqref{rLRVM}.

Despite the mentioned DDEs have been analyzed against the wealth of  cosmological observations (including the CMB data), as indicated in the above  list of references, in the present paper we will  focus our analysis exclusively on  observational data of cosmographic  type, such as the Hubble diagram of distant supernovae (SnIa), data on baryonic acoustic oscillations (BAOs) and also on higher redshift  observations of quasars (QSOs) and gamma-ray bursts (GRBs).  A variant of dynamical DE models not pertaining to the RVM class will also be tested in this work  so as to provide a better comparison among the different DDE frameworks in the light of the cosmographic data. These are the so-called `Ghost DE models' (GDEs) (see e.g.  \cite{Urban:2009vy,Urban:2009yg,Ohta:2010in,Zhitnitsky:2011tr} and references therein). These models are conceptually very different from the RVMs.   It is nonetheless interesting to compare the performance of the GDEs versus the RVMs using the same set of cosmographic data  since all these models predict a dynamical character for the DE associated to powers of $H$.

The idea of using cosmography is to make the analysis as model-independent as possible.  The cosmographic  formulation proceeds by making a  minimal number of  assumptions on the expansion dynamics, namely, it does not assume any particular form of the Friedmann equations.  It only relies on the assumption that the spacetime geometry is well described on large scales by the homogeneous and isotropic  Friedmann-Lema\^\i tre-Robertson-Walker   (FLRW) metric.  The model dependence enters  only in the late stages of interpretation of the results where
one has to borrow the theoretical framework in which the conclusions fit best (say, dark energy, modified gravity, etc.).
The cosmographic analysis should help also to break degeneracies observed in other types of analyses. The method has been used e.g. in\,\citep{Visser:2003vq} to analyze the equation of state of the cosmic fluid in a model-independent way.
The basic procedure in all cases stems from the Taylor series expansions of the scale factor $a(t) $ and related quantities (the so-called cosmographic parameters). The precise methodology is exposed in  Sect. \ref{sect:cosmography}.

The cosmographic approach as a way to distinguish between different DE scenarios was proposed in \citep{Alam:2003sc}, where a variety of models of DE were analyzed using the statefinder parameters\,\citep{Sahni:2002fz}. They study e.g. how to distinguish  a cosmological constant from quintessence models, Chaplygin gas and braneworld models.  In \citep{Capozziello:2009ka}  it was particularly analyzed  how $f(R)-$gravity could be useful to solve the problem of a mass profile and dynamics of galaxy clusters. In \citep{Aviles:2012ir} the $f(R)$ models  were further explored and constrained within the cosmographic method and confirmed that they can pass the cosmographic tests and reproduce the late time acceleration in agreement with observations. The results of \citep{Capozziello:2011tj}, for instance,  show a considerable deviation from the $\Lambda$CDM cosmology by using the cosmographic method. Worth noticing are also the recent works \citep{Gomez-Valent:2018hwc,Gomez-Valent:2018gvm}, where  it is demonstrated, among other things,  that on pure cosmographic grounds the confidence level at which we may claim
that the Universe  is currently accelerating is moderate  if we rely on the data on SnIa+cosmic chronometers (CCH) only, whereas it is very strong  when  data on BAOs is also used.  In \citep{Capozziello:2018jya}, using the low-redshift data sets of cosmography, the authors confirmed the tensions with $\Lambda$CDM model at low redshift universe. From the cosmographic perspective the authors of \citep{Li:2019qic} contend that the most favored cosmographic model prefers a curved universe with $\Omega_K=0.21\pm0.22$. Furthermore, in \citep{Rezaei:2020lfy}, it is shown  that the $\Lambda$ cosmology has serious issues when  quasar and gamma-ray burst data are involved in the analysis; and that while high-redshift QSOs and GRBs can falsify the concordance model, the other DE parameterizations, CPL and PADE, are still consistent with these observations.

 Following the above studies, in this work we focus on the DDEs and in particular on the RVMs  in the context of a strict cosmographic point of view and verify if their success with the cosmological data is reinforced also within the strict cosmographic approach.  At the same time we study the GDEs as an alternative sort of DDEs amply considered in the literature and which are relatively close to the RVMs, although  conceptually very distinct.   We compare these models with a simultaneous cosmographic analysis of the standard $\Lambda$CDM.  On using the Markov Chain Monte Carlo (MCMC) method\footnote{\jt{In order to find the minimum value of least square function, $\chi^2$, we use the Metropolis-Hastings condition in the MCMC algorithm\,\citep{Metropolis,Hastings}.}} and by combining different data sets including the Pantheon collection of SnIa, the BAO sample, the distance modulus of QSO and the GRB data, we first obtain the best-fit values of the cosmographic parameters in a model independent way. Then using the same data samples we put constraints on the free parameters of the DDEs along with the  $\Lambda$CDM. Assuming the  best fit parameters thus obtained we calculate the cosmographic parameters for the models under study.  Finally we compare the cosmographic parameters obtained for each model with those we obtain from the model independent approach. Using different combinations of cosmographic data sets and applying well-known powerful information criteria, we assess the cosmographic performance of the various DDEs versus the $\Lambda$CDM.  We find that the RVMs and GDEs respond in a very different way to their confrontation with the cosmographic data,  showing that this method can be very useful to discriminate between these two classes of DDEs.  All in all  we find it very useful to perform a comparative cosmographic study of the DDE models, and in particular the RVM ones which hint at interesting connections between cosmology and fundamental physics.

 The paper is organized as follows:  In Sect.\, \ref{sect:cosmography}, we review the cosmographic formalism. Subsequently we describe the observational data sets which we have used in our analysis. In Sect.\,\ref{sect:DDE}  we  introduce the various DDE models  and  particularly the RVM  subclass.   In Sect.\,\ref{sec:numerical} we provide the numerical results which we have obtained for the different DE models  and discuss  these results. \jt{We devote Sect.\,\ref{sec:truncation} to estimate the  truncation error  in the Taylor expansion used in our cosmographic approach.}  We \jt{complete the discussion of the model comparison  in terms of different model selection criteria} in Sec.\,\ref{sec:selection}.  Finally,  in Sect.\,\ref{conlusion}  we deliver our main conclusions.

\section{cosmographic approach and data sets}\label{sect:cosmography}

Cosmography is a practical approach to cosmology which helps us to obtain  information from observations mostly in a model independent way.  With such a method one can investigate the evolution of the Universe by assuming just the minimal priors on isotropy and homogeneity.  By adopting the Taylor expansions of the basic observables,  starting from the scale factor $a(t)$, we can extract useful information about the cosmic flow and its evolution \citep{Demianski:2016dsa}. Assuming only the FLRW metric we can obtain the distance - redshift relations from these Taylor expansions. Therefore one could have, in principle,  fully model independent relations. Using first derivatives of the scale factor $a(t)$, we define the cosmographic functions as e.g. in \citep{Capozziello:2011tj}:

\begin{eqnarray}\label{eq1}
\text{Hubble function:}~~& H(t)=\frac{1}{a}\frac{da}{dt}\;,
\end{eqnarray}

\begin{eqnarray}\label{eq2}
\text{deceleration function:}~~& q(t)=-\frac{1}{aH^2}\frac{d^2a}{dt^2}\;,
\end{eqnarray}

\begin{eqnarray}\label{eq3}
\text{jerk function:}~~& j(t)=\frac{1}{aH^3}\frac{d^3a}{dt^3}\;,
\end{eqnarray}

\begin{eqnarray}\label{eq4}
\text{snap function:}~~& s(t)=\frac{1}{aH^4}\frac{d^4a}{dt^4}\;,
\end{eqnarray}

\begin{eqnarray}\label{eq5}
\text{lerk function:}~~& l(t)=\frac{1}{aH^5}\frac{d^5a}{dt^5}\;.
\end{eqnarray}
Except for  the Hubble parameter,  Eq.\,\eqref{eq1}, which has dimension $+1$ in natural units, the other four cosmographic parameters are dimensionless. The names of the dimensionless ones are inspired from the corresponding terminology applied  to the first derivatives of the position function in kinematical studies.  They are related to  the statefinder parameters \citep{Sahni:2002fz,Alam:2003sc}, sometimes partially sharing the notation (although these authors use the notation $r$ for jerk as exactly defined here, but their $s$ is not the snap defined above, for example).
Setting $a=1$ in the above functions we  find the current values of the cosmographic parameters: ($H_0,q_0,j_0,s_0, l_0$). By using the relation between the derivatives of $H$ and the cosmographic parameters, we may compute the Taylor series expansion of the Hubble parameter, up to the forth order in redshift $z$ around $z = 0$ \citep[for detailed disscusion we refer the reader to][]{Rezaei:2020lfy}:
\begin{eqnarray}\label{hexpandz}
H(z)& \simeq &H\vert_{z=0}+\left.\frac{dH}{dz}\right|_{z=0} \dfrac{z}{1!}+\left.\frac{d^2H}{dz^2}\right|_{z=0} \dfrac{z^2}{2!}  \nonumber\\
&+&\left.\frac{d^3H}{dz^3}\right|_{z=0} \dfrac{z^3}{3!}+\left.\frac{d^4H}{dz^4}\right|_{z=0} \dfrac{z^4}{4!}\;.
\end{eqnarray}
The above expansion is valid only for $z <1$, while some of the most interesting observations occur at $z>1$. In other words, the radius of convergence of a series expansion in $z$ is in the redshift interval $(z\lesssim 1)$, therefore a $z$-based expansion will break down for any $z > 1$.  In this work, for instance,  we use
different data sets in the redshift range up to $z=6.67$.  %which force us to ignore the Taylor expansion.
An improved redshift definition is commonly used in literature, which can solve this problem; the `$y$-redshift', defined as $y=\frac{z}{z+1}$ \citep{Cattoen:2007sk}.  Trading  the $z$-redshift for the $y$-redshift not only does not change the physics,  it can  improve significantly the convergence of the cosmographic series expansions.  In terms of the y-redshift, a Taylor series expansion convergerges in the compact interval $0\leq y\lesssim 1$ or, equivalently, in the unbounded interval $0\leq z<\infty$ of the original redshift. Thus, using $y$-redshift, we can apply the Taylor expansion of the Hubble rate at higher redshifts as follows:
\begin{eqnarray}\label{hexpandy}
H(y) \simeq H\vert_{y=0}+\left.\frac{dH}{dy}\right|_{y=0} \dfrac{y}{1!}+\left.\frac{d^2H}{dy^2}\right|_{y=0} \dfrac{y^2}{2!}  \nonumber\\
+\left.\frac{d^3H}{dy^3}\right|_{y=0} \dfrac{y^3}{3!}+\left.\frac{d^4H}{dy^4}\right|_{y=0} \dfrac{y^4}{4!}\;.
\end{eqnarray}
We note that there are some other alternatives which can solve the convergence problem \citep[see also][]{2020MNRAS.494.2576C}.
On the other hand, we are not using the high-redshift CMB data, thus we can apply an approach which performs well at low and intermediate redshifts. Assuming this limitations, and to prevent complexity arising from inserting more additional degrees of freedom, here we choose the $y$-redshift formulation.
Assuming  Eq.(\ref{hexpandy}), the evolution of $H(y)$ or, equivalently, the evolution of $E(y)=\dfrac{H(y)}{H(y=0)}$, we can study the dynamical cosmic fluid in a model independent approach. One  can easily obtain the time derivatives of $H$ with respect to $y$ and insert the results in Eq.(\ref{hexpandy}).  Notice that
\begin{equation}\label{eq:changevariables}
 \dfrac{d}{dt}=-(1+z)H \dfrac{d}{dz}=-\dfrac{H}{1+z}\dfrac{d}{dy}\,,
\end{equation}
and from here one can also compute the higher order derivatives, which are necessary to account for the cosmographic parameters Eqs.\,(\ref{eq1}-\ref{eq5}). The final result follows after a somewhat lengthy but straightforward calculation, which can be cast as follows:
\begin{eqnarray}\label{ey}
E(y) =\dfrac{H(y)}{H(0)} \simeq 1+A y+\dfrac{B y^2}{2}+\dfrac{C y^3}{6}+\dfrac{D y^4}{24}\;,
\end{eqnarray}
where the coefficients $A,  B, C$ and $D$ can be expressed in terms of the current values of the cosmographic parameters as follows:
\begin{eqnarray}\label{k1}
A=1+q_0\;,
\end{eqnarray}
\begin{eqnarray}\label{k2}
B=2-q^2_0+2q_0+j_0\;,
\end{eqnarray}
\begin{eqnarray}\label{k3}
C=6+3q^3_0-3q^2_0+6q_0-4q_0j_0+3j_0-s_0\;,
\end{eqnarray}
\begin{eqnarray}\label{k4}
&&D=-15q^4_0+12q^3_0+25q^2_0j_0+7q_0s_0-4j^2_0-16q_0j_0\nonumber \\
&&\phantom{xxx}-12q^2_0+l_0-4s_0+12j_0+24q_0+24\;.~~
\end{eqnarray}
In order to constrain the cosmographic parameter space using the Hubble diagram of low redshift data, we take the cosmographic parameters ($q_0,j_0,s_0$ and $l_0$) as the free parameters in the MCMC algorithm. Then, the best fit values for these parameters are those which can minimize the $\chi^2$ function. Notice that the definition of  $\chi^2$  is  based on the distance modulus of the observational objects. Hence  using Eq.(\ref{ey}), we first calculate the luminosity distance and then the distance modulus. In parallel way, one can use the logarithmic expansion of the luminosity distance in terms of redshift \citep{Risaliti:2018reu,Bargiacchi:2021fow,Lusso:2019akb}.

The Hubble diagrams corresponding to the  low-redshift data sets used in our cosmographic analysis  are the following:

\begin{itemize}
\item Pantheon sample of SnIa:  It comprises  the set of the latest 1048 data points for the apparent magnitude of type Ia supernovae of \citep{Scolnic:2017caz} in the range $0.01 < z < 2.26$;
%is one of three sample of data points we use in this work.

\item BAOs:  For the baryonic acoustic oscillations sample we consider the radial component of the anisotropic BAOs obtained from measurement of the power spectrum and bispectrum from BOSS data release 12 galaxies \citep{Gil-Marin:2016wya}, the complete SDSS III $Ly\alpha$-quasar \citep{duMasdesBourboux:2017mrl} and the SDSS-IV extended BOSS data release 14 quasar sample \citep{Gil-Marin:2018cgo}.

\item Gamma ray bursts (GRBs): This sample is a compilation of 162 data points in the range $ 0.03 < z < 6.67$. These data points are constructed by calibrating the correlation between the isotropic equivalent radiated energy and the peak photon energy \citep{Amati:2008hq}. For a detailed discussion on this sample, and in particular the  calibration of the distance modulus-redshift relation, we refer the reader to \citep{Demianski:2016zxi,Demianski:2016dsa,Amati:2013sca}.

\item Quasars (QSOs): There is a tight non-linear relation between the X-ray and UV emission in QSOs. Such a  non-linearity leads to a new powerful way  to estimate the absolute luminosity. This capability turns quasars into a new class of standard candles \citep{Lusso:2017hgz}. The main QSOs sample is composed of 1598 data points in the range $0.04<z<5.1$. \jt{ In this work, we follow the approach used in \citep{Risaliti:2015zla} where the large quasar sample has been divided in to 25 redshift bins. The size of each bin is chosen in such a way that $\Delta[\log D_L]<0.10-0.15$ where $D_L$ is the luminosity distance. This condition lets us to consider the maximum size of the redshift bins as $\Delta[\log z]<0.1$. By these considerations, each bin contains enough quasars sources to test the redshift independency of the $F_X-F_{UV}$ relation which is required for quasar observations as standard candles \citep{Risaliti:2015zla,Bisogni2018}.}

\end{itemize}

Using different combinations of Pantheon SnIa data points, as well as BAO, QSO and GRB data, we can probe a wide redshift range $( 0.01 < z < 6.67)$ which is most suited for studying DE physics. Using different combinations of datasets we calculate the $\chi^2$ function of the distance modulus by using the MCMC algorithm. Following this procedure we may determine the best fit values of the cosmographic parameters  $q_0,j_0,s_0$ and $l_0$  in an essentially model independent fashion.

\section{Dynamical DE models:  RVM\lowercase{s} and GDE\lowercase{s}}\label{sect:DDE}

In this work, we focus on two scenarios for dynamical dark energy (DDE) whose energy density can be expressed as a power series expansion of the Hubble rate (and its time derivatives):  $\rho_{\rm de}=\rho_{\rm de}(H,\dot{H},...)$.  These will be generically called the dynamical vacuum models (DDEs).  The DDEs can be of different types,  but for definiteness we focus here on the following two:  i) ghost DE models (cf. \cite{Urban:2009vy,Urban:2009yg,Ohta:2010in,Zhitnitsky:2011tr}), and   ii)  `running vacuum models' (RVMs) as given in Eq.\,\eqref{rLRVM}.  The RVMs are particularly interesting since they can be linked to quantum corrections of the effective action in  QFT in curved spacetime, in particular to the renormalization group \citep{Shapiro:2000dz,Sola:2007sv}, see \citep{Sola:2013gha,Sola:2015rra} for a review and  references therein. In addition,  specific QFT calculations in a flat three-dimensional FLRW background have recently demonstrated that the structure of these models can be explicitly derived from first QFT principles (cf. \,\cite{Moreno-Pulido:2020anb}).  One further step along this promising direction,  worth being remark, is  that the structure of the RVMs can also be motivated from the low-energy effective action of string theory, see \citep{Mavromatos:2020kzj}  for a review and references.  Thus, the RVMs are theoretically well-motivated from different perspectives. At the present epoch, the relevant running terms of the power series expansion in Eq.\,\eqref{rLRVM} can only be of order $H^2$ at most (this includes $\dot{H}$ since it has the same dimension as $H^2$), whereas the higher order terms  $H^{2n} (n>1)$ and associate derivatives of equal dimensionality  can be used in the early Universe to successfully implement inflation, see e.g.\,\citep{Lima:2012mu} and the extensive study \cite{SolaPeracaula:2019kfm}.

We note that despite the name running  vacuum, the equation of state (EoS) for the D3 and D4 models  need not be exactly equal to  $-1$, but in any case it has to be close to $-1$ at present.  The situation when the EoS  can remain exactly equal to $-1$  (hence  being strict  vacuum all the time) occurs e.g.  when the DDE  is interacting with matter, as previously studied in\,\citep{Gomez-Valent:2014rxa,Gomez-Valent:2014fda,Sola:2015wwa} and have been further investigated in \citep{Sola:2016jky,Sola:2017jbl}, see \citep{Sola:2016zeg} for a review.  But we need not restrict the EoS to be of pure vacuum type. If we allow for matter to be self-conserved, and hence the  DDE remaining  also self-conserved, in that case we must have a nontrivial EoS for the DDE which evolves with the cosmological expansion (cf. \cite{Gomez-Valent:2015pia}). We shall address precisely this situation here and in this way we can compare the outcome of the current cosmographical analysis with our previous study of the same models using different cosmological data -- which included CMB,  BAOs and structure formation data (LSS), see\,\citep{Rezaei:2019xwo}.

Models in the first subclass of DDE under study, i.e.  the ``ghost DE models'' (GDEs),  do  {\emph not} include a constant additive term in the $\rho_{\rm de}=\rho_{\rm de}(H,\dot{H},...)$  series  and only  involve powers of the Hubble term with different dimensions, such as $H$ and/or $H^2$.  These models are related to QCD, in an attempt to account for the DE nature within the context of strong interactions.  The important distinction in the case of the RVMs is that they include  a non vanishing constant additive term ($c_0\neq 0)$, as indicated in Eq.\,\eqref{rLRVM}\,\citep{Sola:2013gha,Sola:2015rra}.  Only those containing such a non vanishing additive constant have a smooth limit connection with the  $\CC$CDM.   We will consider two models ($D1$ and $D2$) in the first subclass and two models ($D3$ and $D4$) in the second subclass.  More specifically,  the DDEs addressed in this study read as follows:
\begin{itemize}
\item  Ghost dark energy models (GDEs). Their DE density is linear in  $H$ or  contains  also a power $H^2$:
\begin{eqnarray}\label{gde1}
D1: \rho_{\rm de}(z)&=&\alpha\, H(z)\;\\
D2: \rho_{\rm de}(z)&=&\alpha H(z)+ \beta H^2(z).\label{gde2}
\end{eqnarray}
These models are inspired in the Veneziano ghost field model in QCD theory\,\citep{Veneziano:1979ec} and have been discussed in several phenomenological DE studies, see e.g. the aforesaid references and also \citep{Cai:2012fq}  and references therein.  The ghosts are required to exist for the resolution of the $U_A(1)$ problem\,\citep{Veneziano:1979ec}, but are completely decoupled from the physical sector, which is why they are called ghosts.  While the latter are decoupled from the physical states and make no contribution in the flat Minkowski space, once they are in an expanding background  the cancellation of their contribution
to the vacuum energy is off-set, leaving a small energy density of order $ \Lambda^3_{QCD} H\sim \left(10^{-3} {\rm eV}\right)^4$, which is of the order of the vacuum energy density at present,  see e.g. \citep{Urban:2009vy,Urban:2009yg,Ohta:2010in,Zhitnitsky:2011tr}.
 The D2 model above is a generalization of D1 and is motivated by the fact that the vacuum energy of the Veneziano ghost field introduced to solve the $U(1)_A$ problem in
QCD is of the form  $ \rho_{\rm de}\sim H + {\cal O} (H^2)$.   Notice that the $H^2$ component in Model D2 is not necessarily suppressed as compared to the linear term in $H$,  as the coefficient $\alpha$ and $\beta$ have different natural dimensions of energy ($+3$ and   $+2$ respectively). In the GDE context, coefficient $\alpha$ is expected to be of order of the third power of the QCD scale ( $\Lambda^3_{QCD}$), whereas  $\beta$ is expected of order $m_{Pl}^2=1/G$, with $G$ being Newton's constant. Here $m_{Pl}\sim 10^{19}$ GeV is the usual Planck mass. The ratio of the two coefficients, $\alpha/\beta\sim \Lambda^3_{QCD}/m_{Pl}^2$,  is roughly of the order of the current value  of the Hubble parameter:  $ H_0\sim 10^{-42}$ GeV. It follows that the two terms  in $D2$ are both of order $m_{Pl}^2H_0^2$ for the present universe, and hence  are close to the present value of the vacuum energy density, $\rho_{\Lambda 0}\sim 10^{-47}$ GeV$^4$. This is of course why the GDEs were proposed in the literature, in an attempt to link the value of the cosmological constant to the scales of QCD and Planck. A criticism that can be made to these QCD models of the DE  is that it is very difficult to understand how the vacuum energy density (which should emerge from the covariant effective action in curved spacetime) can be linear in the Hubble rate (for more detailed discussions, see e.g. \citep{Sola:2013gha,Sola:2015rra}).  Even so it is worth trying to check the behavior of these models in the phenomenological context. And this is done here in combination with the RVM models. Let us point out that for convenience in the numerical analysis we will trade the parameter $\beta$ of model D2 for the new parameter $\gamma$, defined as $\gamma=1-8\pi G\beta/3$, similarly as in our previous work \citep{Rezaei:2019xwo}.

\item Running vacuum models (RVMs). As previously discussed, they have a theoretically well motivated status\,\citep{Sola:2013gha,Sola:2016zeg}. The variant that we will analyze here (characterized by self-conservation of matter and vacuum energy) was proposed in \citep{Gomez-Valent:2015pia} and further analyzed in \citep{Rezaei:2019xwo}.
The basic two types of RVMs  under consideration which are relevant for the cosmographical analysis can be obtained from Eq.\,\eqref{rLRVM} by neglecting the higher order powers of $H$. They can be cast as follows:
\begin{eqnarray}\label{gde3}
D3: \rho_{\rm de}(z)&=&\dfrac{3}{\kappa^2}\left[ c_0+\nu H^2(z)\right] \,,\\
D4: \rho_{\rm de}(z)&=&\dfrac{3}{\kappa^2}\left[c_0+\dfrac{2}{3}\mu \dot{H}(z)+\nu H^2(z)\right]\,.\label{gde6}
\end{eqnarray}
Here we have redefined the parameter $\tilde{\nu}$ of Eq.\,\eqref{rLRVM}  as  $\tilde{\nu}\equiv\dfrac{2}{3}\mu$ just for convenience and also for better comparison with our previous work \citep{Rezaei:2019xwo}.  Let us note that despite $D3$ is a particular case of $D4$, the former has one parameter less and is the canonical RVM\,\citep{Sola:2013gha,Sola:2016zeg}.
In the above equations, the  parameter $c_0$ has dimension $+2$ in natural units, see below.  It is important to note that  for $\nu,\mu\to 0$ these two RVMs  reduce to the $\CC$CDM, in contrast to the GDEs, D1 and D2.  In actual fact, $\nu$ and $\mu$ must remain both small in absolute value, in contrast to the parameters $\alpha$ and $\beta$ of \eqref{gde1} and \eqref{gde2}.  This is because the RVMs, in contrast to the GDEs, remain naturally close to the $\CC$CDM model.  The above two models  can be realized both as strict vacuum models or with a nontrivial EoS slightly departing from $-1$. We choose the last option here and for this to be so we need that the DDE density is covariantly  self-conserved, viz. $\dot{\rho}_{de} + 3H(\rho_{de}+p_{de}) = 0$, independently of the  matter and radiation  self-conservation equations.  Therefore, for the current work we adopt the local conservation laws:

 \begin{eqnarray}\label{continuity}
 && \dot{\rho_{\rm r}}+4H\rho_{\rm r}=0\;,\label{radiation}\\
&&\dot{\rho_{\rm m}}+3H\rho_{\rm m}=0\;,\label{matter}\\
&&\dot{\rho_{\rm de}}+3H(1+w_{\rm d})\rho_{\rm de}=0\;\label{de},
 \end{eqnarray}
where $w_{\rm d}=p_{de}/\rho_{\rm de}$ is the nontrivial EoS, which can be determined upon solving the models. In point of fact the above set of conservation laws actually holds good  for all the DDEs studied here, both the RVMs and GDEs.
In this sense the comparison with the two sorts of models can be made more leveled.  We emphasize that the same RVMs can be studied under the assumption of exact vacuum EoS ($w=-1$) but only at the expense of assuming interaction with matter, see \citep{Sola:2016zeg} for a review, but we shall not adopt this point of view here  but rather adhere to  the same framework we considered before in \,\citep{Rezaei:2019xwo}.
\end{itemize}

\begin{table*}
 \centering
 \caption{The best fit values of the cosmographic parameters and their $1\sigma$ uncertainties obtained for different data combinations in the model-independent (MI)  approach.
}
\begin{tabular}{c  c  c c c}
\hline \hline
Data sample & $q_0$ & $j_0$ & $s_0$& $l_0$ \\
\hline
Pantheon & $-0.702\pm 0.104$ &$1.60\pm 0.71$ &$-3.54^{+0.38}_{-1.5}$ &$-4.9^{+6.3}_{-5.0}$ \\
\hline
Pantheon+BAOs & $-0.609\pm 0.094$ &$0.892\pm 0.44$ &$-2.98^{+0.42}_{-1.02}$ &$-4.4^{+3.3}_{-4.2}$ \\
\hline
Pantheon+QSOs  &$-0.844\pm 0.048$ &$ 2.42\pm 0.25$ &$-2.5^{+1.4}_{-1.2}$ &$-3.2^{+2.5}_{-2.1}$  \\
\hline
Pantheon+QSOs +GRBs  &$-0.819\pm 0.065$ &$ 2.21^{+0.37}_{-0.42}$ &$-3.44^{+0.46}_{-1.5}$ & $-3.8^{+8.2}_{-6.2}$ \\
\hline \hline
\end{tabular}\label{tab:bestmi}
\end{table*}
\begin{figure*}
	\centering
	\includegraphics[width=11cm]{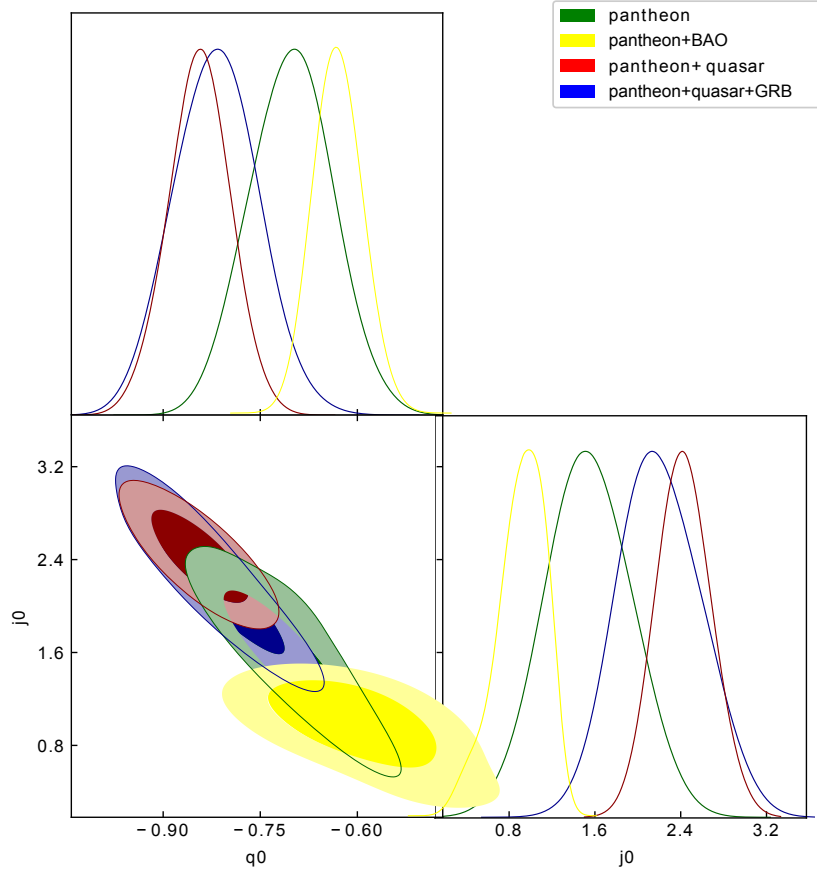}
		\caption{The $1\sigma$ and $2\sigma$ confidence regions in the $q_0 - j_0$ plane obtained from the MI approach from different combinations of data sets.}
	\label{fig:back}
\end{figure*}

In order to establish useful relations among the cosmographic parameters in DDE cosmology,  let us note that Eq.\eqref{eq2} can be rewritten as follows:
\begin{eqnarray}
q=-\dfrac{a\ddot{a}}{\dot{a}^2}=-1-\dfrac{\dot{H}}{H^2}\;.\label{qqq}
 \end{eqnarray}
At the same time it is easy to derive
\begin{eqnarray}
\dfrac{\dot{H}}{H^2}=-\dfrac{3}{2}(1+w_{\rm d}\Omega_{\rm d})\;.\label{qq}
\end{eqnarray}
Upon inserting the values of $w_{\rm d}$ and $\Omega_{\rm d}$ for each of the DDEs in Eq.\eqref{qq} and replacing the result in Eq.\eqref{qqq}, we can obtain the explicit form of the deceleration parameter $q$ for  the different models. For the other cosmographic parameters, using equations\,\eqref{eq1}-\eqref{eq5} and \eqref{qqq},  we reach the useful formula \citep{Aviles:2012ir}:
\begin{eqnarray}
j=-\dfrac{dq}{d \ln a}+2(1+q)^2-3q-2\,.\label{j}
\end{eqnarray}
Furthermore in the case of the snap and the lerk parameters, $s$ and  $l$, one obtains
\begin{eqnarray}
s=\dfrac{dj}{d \ln a}-2j-3qj\label{s}
\end{eqnarray}
and
\begin{eqnarray}
l=\dfrac{ds}{d \ln a}-3s-4qs\,.\label{l}
\end{eqnarray}
 As for the various EoS of the DDEs,  they were  considered in our previous study \citep{Rezaei:2019xwo}. Let us only quote here the result for Model D4.  Expanding the exact solution  for small redshift $z$, i.e. around our current epoch,  the effective EoS for such a model is
\begin{eqnarray}
\label{eq:EoSmod3}
w_{\rm de}(z)&\simeq&-1+\frac{H_0^2(1-\nu)}{c_0}\,\Omega_{m,0}\,(\nu-\mu)\,(1+z)^3\nonumber\\
&\simeq& -1+\frac{\Omega_{m,0}}{1-\Omega_{m,0}}\,(\nu-\mu)\,(1+z)^3\,,
\end{eqnarray}
where $c_0$ obtains from matching the current value of the DE density of model D4  to the value of the cosmological constant. Its exact expression is given by
\begin{equation}\label{35}
c_0=H^2_0[\Omega_{\rm de,0}-\nu+\mu(1+w_{\rm de,0}\Omega_{\rm de,0}+\dfrac{\Omega_{\rm r,0}}{3})]\;.
\end{equation}
In the expansion \eqref{eq:EoSmod3} we have neglected terms beyond linear order in $\nu$ and $\mu$ in the final result since these are small for Model D4 (and a fortiori for Model D3).  Needless to say, the corresponding EoS for Model D3 is obtained from the previous formulas by setting $\mu=0$. Besides,  for $\mu,\nu\to 0$ we recover the EoS of the $\CC$CDM, i.e. $w_{\rm de}\to -1$. This last feature makes models D3 and D4 particularly interesing, a feature which is not shared with models D1 and D2.
Equation \eqref{eq:EoSmod3} is  useful  to assess the effective quintessence or phantom-like behavior exhibited by  models D3 and D4.  To wit: for $\nu-\mu>0$ Model D4 behaves as effective quintessence, whereas for $\nu-\mu>0$ it behaves (effectively) as phantom DE. But there is,  of course,  no real phantom nor quintessence at all here, just an effective behavior.  This also illustrates  how the running vacuum models can mimic exotic cosmological behaviors which in the context of elementary scalar fields could be very problematic with QFT (specifically in the case of phantom DE). In contrast, the RVMs are perfectly well behaved and consistent with QFT in curved spacetime\,\citep{Moreno-Pulido:2020anb}.

\section{Numerical analysis and discussion}\label{sec:numerical}

In this section, we present and discuss our numerical results concerning  the DDEs under study  (D1, D2, D3 \& D4) as defined in the previous section,  including the concordance $\Lambda$CDM cosmology. In the first step of our analysis we obtain the best fit values of the cosmographic parameters in a model-independent way. In order to run the MCMC algorithm,  we  first select the initial values of the free parameters following the method of \citep{Rezaei:2020lfy}, hence we run our MCMC code using two different sets of the initial values.  The free parameters being directly  fitted by the cosmographic data are $(q_0, j_0, s_0, l_0)$ since these are the parameters involved in \eqref{ey}. After running the algorithm we observe that both of the initial sets lead to similar posteriors for $q_0$ and $j_0$ while for the other cosmographic parameters,$s_0$ and $l_0$, the posteriors are slightly different. Thus, we select the initial value for $s_0$ and $l_0$ between the two best-fit values obtained in previous steps. This procedure shows that our results are fully independent from the initial values of the free parameters. Moreover, for each of the free parameters we assume big values of $\sigma$ which can guarantee that the MCMC procedure sweeps the whole of the parameters space. These choices help us to remove the risk of finding local best fit values in the parameters space. Here, we choose the initial value of free parameters similar to those we have chosen in the mentioned previous work, namely  $q_0=-0.8, j_0=2.5, s_0=-2.0, l_0=-3.0$.

To see the effect of different data samples, we assume different combinations of data sets as Pantheon, Pantheon+BAOs, Pantheon+QSOs and Pantheon+QSOs+GRBs. For each of these data combinations, we repeat our analysis to find the best fit parameters and their $1\sigma$ and $2\sigma$ uncertainties. The results of the approach are displayed in Table \ref{tab:bestmi} and  Fig.\,1.  For each of the data combinations, we find that the deceleration parameter $q_0$ is tightly constrained and the constraint on $j_0$ remains still significant, especially for the cases where we combine more data.  We can see that adding high redshift data points of the  QSO and GRB types  leads to higher values of $q_0$ and $j_0$. However, unfortunately,  we can not put tight constraints on the snap $s_0$ and lerk $l_0$ parameters, the errors being larger than $50\%$ for the former and far beyond $100\%$ for the latter, even in the optimal case when we combine  three sources of cosmographic data.  We should notice that the  $s_0$ and $l_0$ parameters start entering the expansion \eqref{ey} only at the level of  the coefficients of the third and forth power of the $y$-redshift, respectively.  For these terms, the bigger error bars of  the data points are responsible for the  weaker constraints on these two parameters. Thus, to make a more efficient comparison between different cosmologies, we focus on the fitting results of $q_0$ and $j_0$ only.

\begin{table*}
  \centering
  \caption{The best fit values of the free parameters obtained for the $\Lambda$CDM model (left part)  and the corresponding computed  values of the cosmographic parameters from equations \eqref{qqq} and  \eqref{j}, shown in the right part . Here $h$ is the reduced Hubble parameter, defined as usual through $H_0=100\, h\, {\rm km}/{\rm s}/{\rm Mpc}$.
 }
 \begin{tabular}{c c c  c c c c  c}
 \hline \hline
  &  best fit parameters& & & $\vert$ & & computed values &   \\
 Data  & $\Omega_{dm,0}$ & $\Omega_{b,0}$ & $h$  & $\vert$ & $q_0$&  &  $j_0$\\
 \hline
 Pantheon & $0.239\pm 0.011$ & $0.045\pm 0.003$ & $0.622\pm 0.031$ & $\vert$ & $-0.573\pm 0.016$ &  & $1.0$\\
 \hline
 Pantheon+BAOs & $0.241\pm 0.013$ &$0.044\pm 0.004$ & $0.691\pm 0.027$ & $\vert$ & $-0.572\pm 0.017$ &  & $1.0$\\
  \hline
 Pantheon+QSOs  & $0.253\pm 0.012$ &  $0.040\pm 0.005$& $0.631\pm 0.027$ & $\vert$ & $-0.561\pm 0.017$ &  & $1.0$\\
 \hline
 Pantheon+QSOs +GRBs & $0.251\pm 0.014$ &$0.044\pm 0.003$ & $0.634\pm 0.033$ & $\vert$ & $-0.557\pm 0.019$ &  & $1.0$\\
 \hline \hline
 \end{tabular}\label{bestlcdm}
 \end{table*}
 %%%%%%%%%%%%%%%%%%%
 \begin{table*}
  \centering
  \caption{The best-fit values of the free parameters for  model D1 (left part) and the corresponding computed  values of the cosmographic parameters  (right part).
 }
 \begin{tabular}{c c  c  c c c  c}
 \hline \hline
  &  & best fit parameters & & $\vert$ & & computed values  \\
 Data & $\Omega_{dm,0}$ & $\Omega_{b,0}$ & $h$ & $\vert$ & $q_0$&  $j_0$\\
 \hline
 Pantheon & $0.1604 \pm 0.029$& $0.003 \pm 0.003$ & $0.648^{+0.020}_{-0.020}$ &  $\vert$ & $-0.526^{+0.064}_{-0.071}$  & $0.021^{+0.47}_{-0.40}$\\
 \hline
 Pantheon+BAOs & $0.1617^{+0.036}_{-0.024}$ &  $0.033^{+0.009}_{-0.006}$ & $0.661^{+0.026}_{-0.023}$ &  $\vert$ & $-0.521^{+0.049}_{-0.051}$  & $2.151 \pm 0.189$\\
  \hline
 Pantheon+QSOs  & $0.169 \pm 0.034$& $0.034^{+0.005}_{-0.004}$ & $0.625^{+0.024}_{-0.021}$ &  $\vert$ & $-0.493^{+0.059}_{-0.062}$  & $0.006^{+0.51}_{-0.42}$\\
 \hline
 Pantheon+QSOs+GRBs & $0.181^{+0.030}_{-0.034}$& $0.037^{+0.003}_{-0.004}$ & $0.633^{+0.022}_{-0.022}$ &  $\vert$ & $-0.463^{+0.061}_{-0.049}$  & $0.034^{+0.37}_{-0.41}$\\
 \hline \hline
 \end{tabular}\label{bestdde1}
 \end{table*}

 \begin{table*}
  \centering
  \caption{As in Table 3,  but  for  model D2.
 }
 \begin{tabular}{c c c  c c c  c  c}
 \hline \hline
  &  & best fit parameters& &  & $\vert$ & & computed  values  \\
 Data & $\Omega_{dm,0}$ & $\Omega_{b,0}$ &$h$ & $ \gamma $ & $\vert$ & $q_0$&    $j_0$\\
 \hline
 Pantheon & $0.152^{+0.033}_{-0.031}$ & $0.031^{+0.006}_{-0.004}$ & $0.647^{+0.019}_{-0.023}$ & $0.854^{+0.019}_{-0.017}$ & $\vert$ & $-0.470^{+0.064}_{-0.067}$  & $0.028^{+0.50}_{-0.51}$\\
 \hline
 Pantheon+BAOs & $0.156^{+0.052}_{-0.049}$ & $0.022^{+0.007}_{-0.003}$& $0.654^{+0.029}_{-0.025}$ & $0.869^{+0.018}_{-0.017}$ & $\vert$ & $-0.507 \pm 0.062$  & $1.831^{+0.29}_{-0.30}$\\
 \hline
 Pantheon+QSOs  & $0.157^{+0.036}_{-0.037}$ & $0.030^{+0.004}_{-0.003}$ & $0.628^{+0.024}_{-0.024}$ & $0.846^{+0.023}_{-0.027}$ &  $\vert$ & $-0.457^{+0.070}_{-0.067}$  & $0.040^{+0.49}_{-0.47}$\\
 \hline
 Pantheon+QSOs+GRBs & $0.169^{+0.034}_{-0.032}$ & $0.032^{+0.006}_{-0.003}$ & $0.603^{+0.027}_{-0.022}$ & $0.851^{+0.021}_{-0.020}$ &  $\vert$ & $-0.426^{+0.062}_{-0.069}$  & $0.063^{+0.52}_{-0.54}$\\
 \hline \hline
 \end{tabular}\label{bestdde2}
 \end{table*}

Now, using the MCMC method on the different dataset combinations, which  include: i) Pantheon, ii) Pantheon + BAOs, iii) Pantheon + QSOs and  iv) Pantheon+QSOs+GRBs, we can determine the best fit values of the free parameters for the various  DDEs as well as their confidence regions at  $1\sigma$ level.
Our results are reported in the left  blocks  of Tables \ref{bestlcdm}-\ref{bestdde4}. Inserting them into equations \eqref{qqq} and  \eqref{j} we may calculate the values of the cosmographic parameters $q_0$ and $j_0$  for the models under consideration, which are each characterized by an specific expression for the corresponding  Hubble function $H(a)$ (cf.  \cite{Rezaei:2019xwo}).  The results for the various DDEs, together with the corresponding  $1\sigma$ confidence regions, are presented on the two rightmost blocks of the mentioned tables.
Let us recall that among the free parameters we have $\Omega_{dm,0}$ (the usual cosmological parameter for the current CDM density),  $ \Omega_{b,0}$ (for the current baryonic density)  and   $h=H_0/(100 \, {\rm km}/s/{\rm Mpc})$  (the reduced Hubble parameter at present). These three parameters enter  the fitting of all the models, including the $\Lambda$CDM.  To these basic free parameters we may have to add the more specific ones carried by each DDE model.  In the D1 case,  however, the parameter $\alpha$ is not free, it is easy to show that $\alpha=\frac{3H_0}{8\pi G} \Omega_{\Lambda, 0}$. For D2 we have $\beta$ (or, alternatively,  $\gamma$,  which was defined in the previous section) as an additionally free parameter;  for  D3 we have $\nu$; and, finally, for D4 we  have both  $\nu$ and $\mu$. However,  in order to break degeneracies between  the parameters of model D4,  in practice we shall  set  $\mu= -\nu$ in the numerical analysis. Since $\nu$ is forced to be negative in this model  -- see \citep{Rezaei:2019xwo}  -- the fitted values of $\mu$ will always be positive in this setting.

 \begin{table*}
  \centering
  \caption{The best-fit values of the free parameters for  model D3 (left part) and the corresponding computed  values of the cosmographic parameters  (right part).
 }
 \begin{tabular}{c  c c c c c  c c c}
 \hline \hline
  &  & best fit parameters & & & $\vert$ & & computed values &  \\
 Data & $\Omega_{dm,0}$ & $\Omega_{b,0}$& $h$ & $\nu $ & $\vert$ & $q_0$&    & $j_0$\\
 \hline
 Pantheon & $0.162 \pm 0.062$ & $0.029^{+0.004}_{-0.002}$ & $0.635^{+0.031}_{-0.036}$ & $0.329^{+0.021}_{-0.019}$ &  $\vert$ & $-0.549^{+0.074}_{-0.070}$ &    & $0.918^{+0.49}_{-0.47}$\\
 \hline
 Pantheon+BAOs & $0.243^{+0.031}_{-0.028}$ & $0.041^{+0.004}_{-0.003}$ & $0.733^{+0.011}_{-0.016}$ & $0.133^{+0.061}_{-0.059}$ &  $\vert$ & $-0.606^{+0.054}_{-0.057}$ &    & $1.389^{+0.23}_{-0.24}$\\
  \hline
 Pantheon+QSOs  & $0.196^{+0.056}_{-0.054}$ & $0.039^{+0.005}_{-0.004}$ & $0.637^{+0.033}_{-0.036}$ & $-0.160^{+0.017}_{-0.020}$ & $\vert$ & $-0.469^{+0.073}_{-0.076}$ &   & $0.778^{+0.64}_{-0.66}$\\
 \hline
 Pantheon+QSOs+GRBs & $0.193^{+0.052}_{-0.051}$ & $0.041^{+0.005}_{-0.002}$ & $0.642^{+0.026}_{-0.022}$ & $-0.270^{+0.013}_{-0.016}$ & $\vert$ & $-0.608^{+0.063}_{-0.062}$ &  & $1.289^{+0.50}_{-0.53}$\\
 \hline \hline
 \end{tabular}\label{bestdde3}
 \end{table*}

 \begin{table*}
  \centering
  \caption{As in Table 5,  but  for  model D4.
 }
 \begin{tabular}{c  c  c c c c  c c c}
 \hline \hline
  &  & best fit parameters & & & $\vert$ & & computed values &  \\
 Data & $\Omega_{dm,0}$ & $\Omega_{b,0}$ & $h$ & $\mu $ & $\vert$ & $q_0$&    & $j_0$\\
 \hline
 Pantheon & $0.171 \pm 0.040$ & $0.034^{+0.005}_{-0.004}$ & $0.656^{+0.047}_{-0.044}$ & $0.098^{+0.022}_{-0.023}$ &  $\vert$ & $-0.637^{+0.071}_{-0.073}$ &    & $3.594^{+0.54}_{-0.56}$\\
 \hline
 Pantheon+BAOs & $0.247 \pm 0.019$  & $0.051^{+0.003}_{-0.004}$ & $0.777^{+0.017}_{-0.032}$ & $0.741^{+0.029}_{-0.036}$ &  $\vert$ & $-0.619 \pm 0.057$ &    & $1.383^{+0.19}_{-0.18}$\\
  \hline
 Pantheon+QSOs & $0.163^{+0.036}_{-0.035}$ & $0.031^{+0.006}_{-0.004}$ & $0.600^{+0.041}_{-0.038}$ & $0.033^{+0.032}_{-0.019}$ & $\vert$ & $-0.661^{+0.062}_{-0.066}$ &   & $5.581^{+0.84}_{-0.89}$\\
 \hline
 Pantheon+QSOs+GRBs & $0.187^{+0.040}_{-0.039}$ & $0.037^{+0.004}_{-0.005}$ & $0.628^{+0.036}_{-0.032}$ & $0.017^{+0.039}_{-0.014}$ & $\vert$ & $-0.534^{+0.073}_{-0.071}$ &  & $5.493^{+0.77}_{-0.72}$\\
 \hline \hline
 \end{tabular}\label{bestdde4}
 \end{table*}

Let us compare the values of the cosmographic parameters derived from the DDEs and the concordance $\Lambda$CDM  with those obtained from the model-independent (hereafter MI)  approach presented in Table \ref{tab:bestmi}. In Fig.\,\ref{fig:back}, we can observe the $1\sigma$ and $2\sigma$ confidence regions of $q_0$ and $j_0$ obtained from the MI method. Quite obviously, the $2\sigma$ confidence region of the jerk parameter lies above the critical value of $j_0=1$ for both Pantheon+QSOs and +QSOs+GRBs, while in the case of the Pantheon+BAOs, the $1\sigma$ confidence region covers $j_0=1.0$ very good.

Recall that,  in the spatially flat $\Lambda$CDM case,  $j=1.0$ is a fixed point of the cosmological evolution\,\citep{Alam:2003sc}. It is instructive to see this by noting that the asymptotic value of the derivative of the total pressure of the cosmic fluid can be related to the jerk. Expanding the total pressure in terms of the total density (i.e. the EoS of the cosmic fluid)  around our epoch,
\begin{equation}\label{eq:expansionp}
p=p_0+\kappa_o (\rho-\rho_0)+{\cal O}(\rho-\rho_0)^2\,,
\end{equation}
one finds that  the first nontrivial coefficient in such an expansion is related to the departure of the cosmological
jerk with respect to the mentioned fixed point:
\begin{equation}\label{eq:kappazero}
\kappa_0=\left.\frac{\dot{p}}{\dot{\rho}}\right|_0=-\frac13\,\frac{1-j_0}{1+q_0}+...
\end{equation}
where the \dots are  terms which are negligible in the limit of small spatial curvature ($|\Omega_0-1|\ll1)$. See \,\citep{Visser:2003vq} for details.
Taking into account that for the $\CC$CDM model $p$ tends to the constant value associated to the cosmological constant ($p\rightarrow p_\CC=-\rho_\CC=$const.), the derivative of $p$ tends to  zero and  thus we have $j_0\rightarrow 1$. This situation is approached at present for the $\CC$CDM, and so if the model of the cosmological evolution is indeed the concordance model, it should necessary follow such a trend (Table 2).  As it turns out, however,  this behavior is not supported by all the dataset combinations used here,  only by the Pantheon and  Pantheon+BAOs samples (cf. Table 1).  In the last two cases the confidence region of $j_0$ covers the critical value  $j_0 = 1$. However,  such a value  is not supported at all by the high redshift cosmographic data associated to QSOs and GRBs. As for the various DDEs,  the output from model D3 stays remarkably close to that of the $\CC$CDM prediction for all datasets. The situation with D4 is nevertheless more peculiar since  it renders  values of $j_0$ well beyond the $\CC$CDM prediction and even beyond the MI  fit for all datasets not involving BAOs  (see Tables 5 and 6).  Notwithstanding, when the BAO data are put back in the fit,  the $j_0$ prediction from model D4  is  driven, too,  towards  the $\CC$CDM prediction.  Therefore, we may conclude  that BAO data are very solid and play a crucial role to stabilize the DE models that are close to the $\CC$CDM. In point of fact, even  the MI fit stays close to the $\CC$CDM prediction in the presence of BAOs, as can be seen from Table 1. In the absence of BAOs,  the concordance  $\Lambda$CDM model (characterized by the fixed point  $j_0=1.0$) as well as  the GDEs (which predict  $j_0\ll1$)  and  also model D4 show all of them significant cosmographic tensions (actually pointing towards disparate directions) with respect to the high redshift data samples used in our analysis, i.e.  QSOs and GRBs.   At the end of the day, the general verdict on the  GDEs is that they  are  in conflict with  the $\CC$CDM and with the MI  fit in all cases,  even if BAOs are kept. This shows that these models of DE  are highly unfavored.   In stark contrast, we  have seen that when the BAO data are present the two RVMs, D3 and D4, predict essentially the same  value  for the jerk parameter. Moreover, such a common value is fairly close to the $\CC$CDM one and is no longer in severe tension with  the MI  value.

\begin{figure*}
	\centering
	\includegraphics[width=8.5 cm]{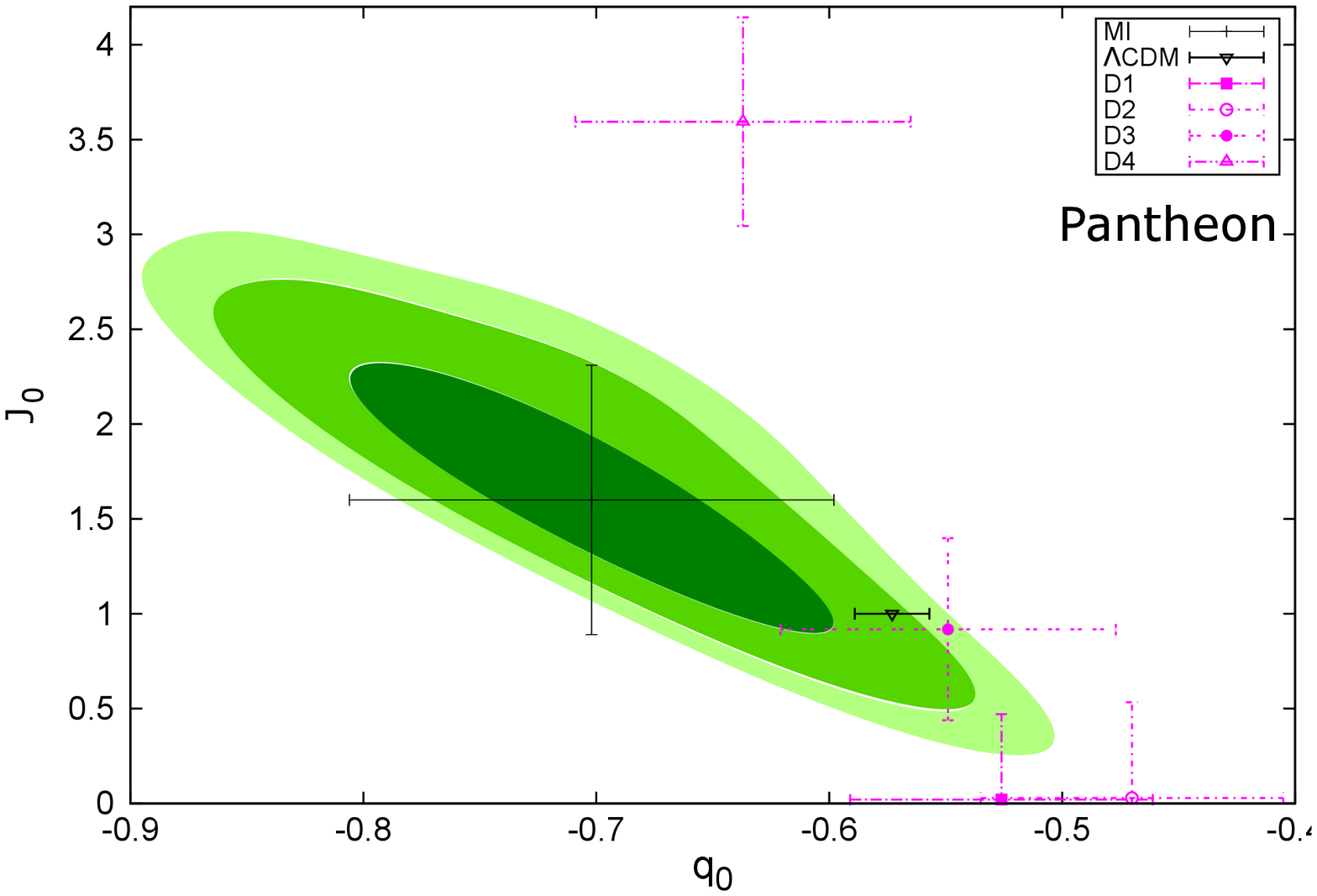}
	\includegraphics[width=8.5 cm]{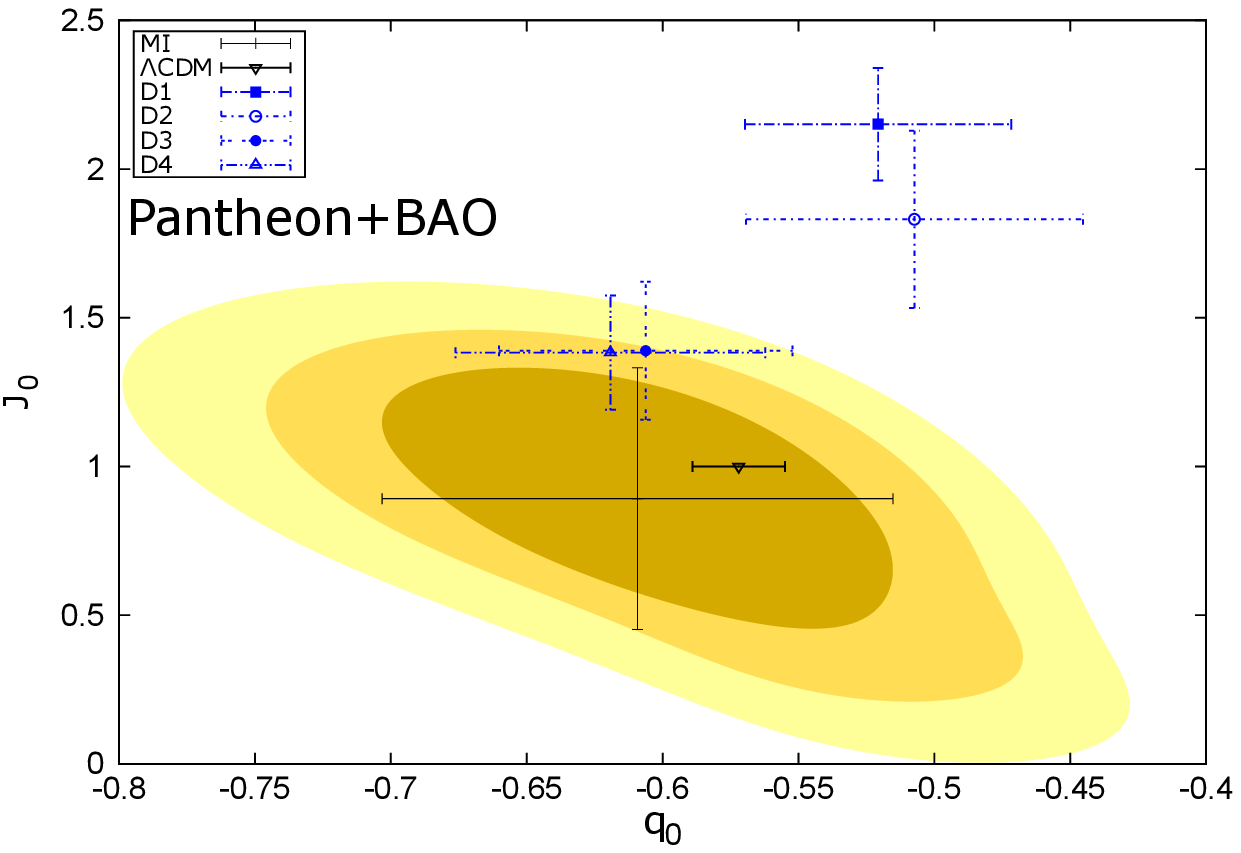}
	\includegraphics[width=8.5 cm]{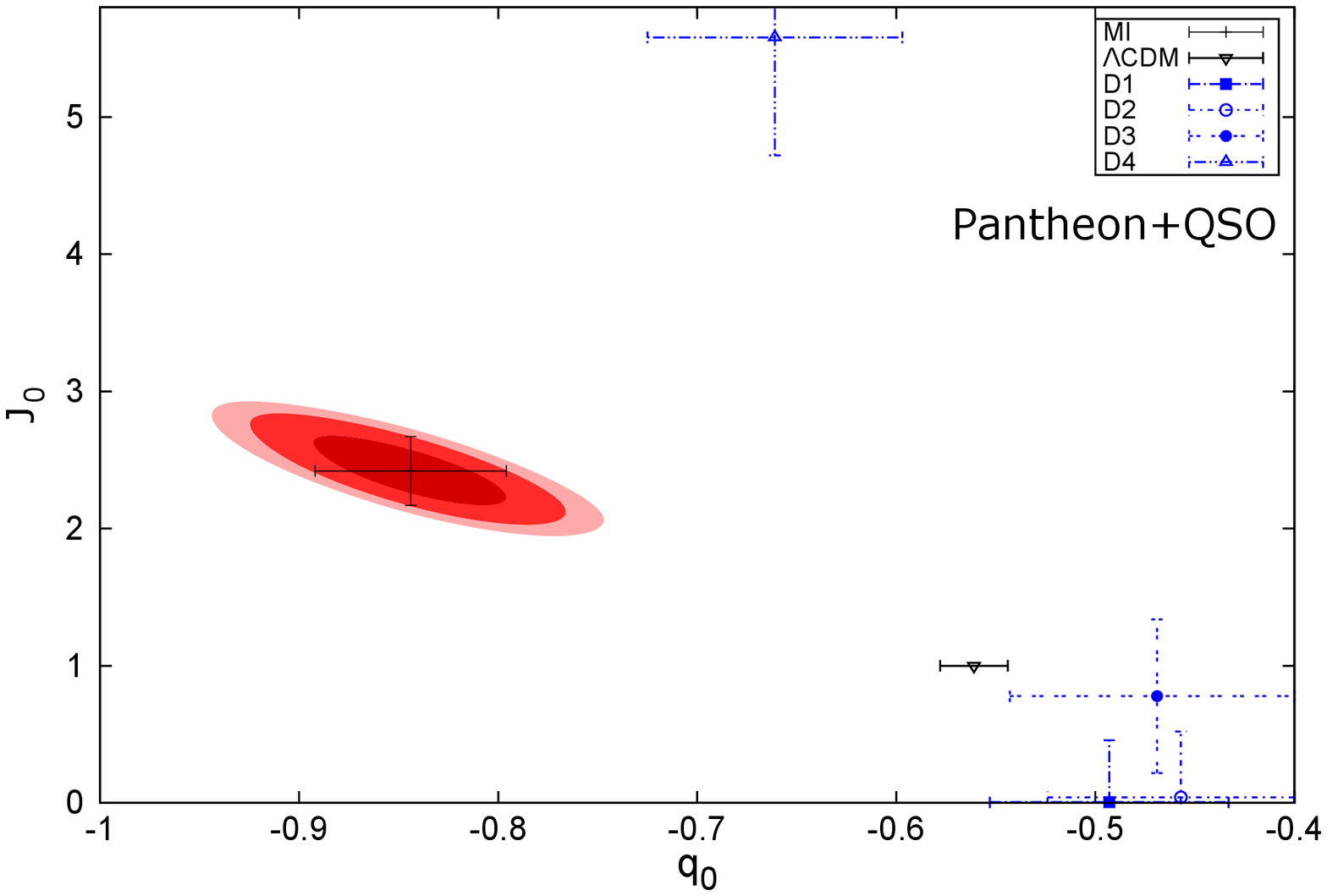}
	\includegraphics[width=8.5 cm]{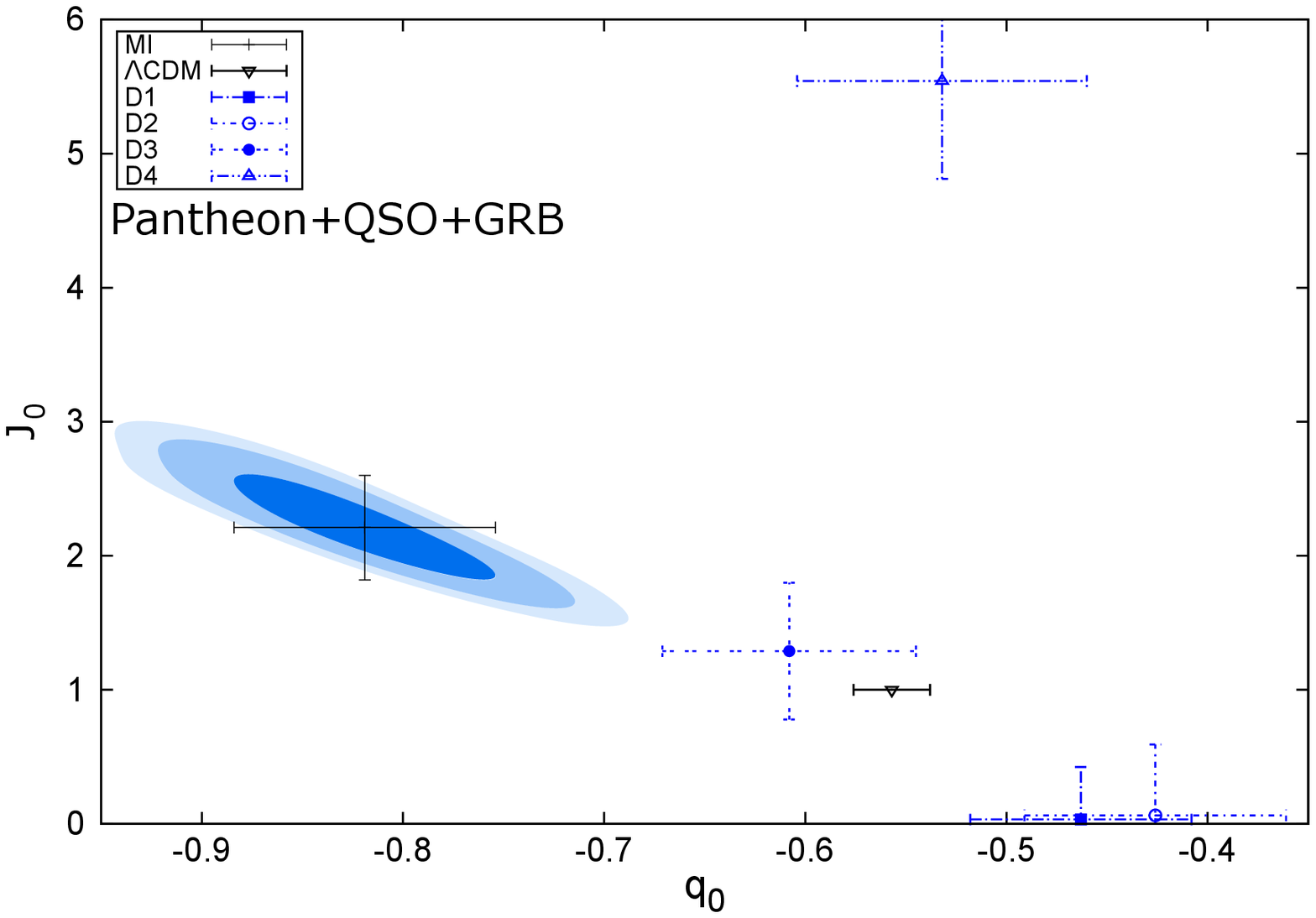}
	\caption{ Contour plots for the confidence levels of the cosmographic parameters $q_0$ and $j_0$ up to $3\sigma$, obtained from the MI  approach. Also shown are the computed values of the same parameters for the different DDE scenarios. The upper-left (right)  panels show the results for the Pantheon (Pantheon+BAOs) data sample combinations. Similarly, the lower-left (right) panels display the corresponding results for the Pantheon+QSOs (Pantheon+QSOs+GRBs) combinations.}
	\label{fig:miandmodels}
\end{figure*}

From the foregoing discussion we can assert  that the quasars and gamma-ray data appear to be more exotic as compared to supernovae and baryonic acoustic oscillations data, in the sense that the former may need some more qualification before we can use them to test cosmological models on equal footing with more consolidated data, such  as  the SnIa+BAO  datasets\,\footnote{For example, in \citep{Khadka:2019njj} it is found that  the QSO and the $H(z)$ + BAO observations are mostly consistent for constraining some cosmological models,  and in \citep{Khadka:2021dsz} the GRBs data are found to exhibit some inconsistencies as compared to  $H(z)$ + BAO data for constraining the cosmological models.  These works, however, do not deal with the cosmographic considerations, as we have done in the current work.}. Interestingly enough, our results concerning the $\CC$CDM can be compared  with those from \citep{Lusso:2019akb}, which confirm the tension between the best fit cosmographic parameters and the $\Lambda$ cosmology at $\sim 4\sigma$ with SnIa+QSO data,  at $\sim 2\sigma$ with SnIa+GRBs and at $> 4\sigma$ with the whole SnIa+QSOs+GRBs data set, see also \cite{Mehrabi:2020zau}. It is therefore not surprising that the RVMs come out with a similarly tensioned behavior with the MI data since these models perform a mild dynamical evolution around the $\CC$CDM, which is however enough to help fitting the overall cosmological data better than the $\CC$CDM\,\citep{Rezaei:2019xwo}.  Our study indeed shows that the response of the DDEs (most particularly the RVM ones, D3 and D4) to the combined Pantheon+BAOs samples is smoother than in the cases when the more exotic QSOs+GRS data are included. We can see e.g.  from Fig. 1 that the response of the Pantheon contours to adding QSOs or QSOs+GRBs is opposite to that of adding BAOs. In the latter case the contours move down and to the right, hence implying smaller values of $j_0$ and smaller (absolute) values of $q_0$, whereas in the former cases the original contour is dragged upwards and to the left, thus leading to  higher values of $j_0$ and larger (absolute)  values of $q_0$.   These features can also be seen in more detail in Fig. 2. As a result the highest values of  $q_0$ and $j_0$  (in magnitude) appear only when the QSOs and/or GRBs are included. This is also reflected, of course,  in the numerical  MI results indicated in Table 1.  As previously noted,  the  RVMs (models D3 and D4)  provide very similar values  of   $q_0$ and $j_0$  only when Pantheon+BAOs data are used (cf. Tables 5 and 6),  whereas $j_0$ becomes much larger for D4 than for D3 when BAOs are not used and are replaced with QSOs+GRBs data.  For this reason we have  not combined BAOs with QSOs and GRBs, as this makes this feature more apparent.

 As for the GDEs (models D1 and D2) their situation appears always precarious, unfortunately,  as they tend to favor lower values of $q_0$, and anomalously low values of  $j_0$, in the presence of QSOs+GRBs. Even though they recover  somewhat when the more robust kind of  BAOs data are used in place of QSOs+GRBs (as can be seen in Tables 3 and 4)  these models appear substantially strayed off against the overall set of observations and go adrift somehow.

Despite the former qualitative considerations, which give already a panoramic view of the results of our analysis,  let  us next discuss the results with some more quantitative detail as to the comparison between  the best fit parameter values of $q_0$ and $j_0$ obtained for each of the DDEs  (see Tables \ref{bestlcdm}-\ref{bestdde4}) and those derived  from the MI approach (see Table\,\ref{tab:bestmi}). For each of the dataset combinations we have the following results:

\begin{enumerate}
\item {\bf Pantheon sample}:   This SnIa sample is certainly a well-established one for testing cosmological models. Within the cosmographical (MI) approach,  we obtain with these data the best fit values for the deceleration and cosmological jerk parameters, together with their $1\sigma$ uncertainties, as follows:  $q_0=-0.702\pm 0.10$ and $j_0=1.60\pm0.71$ . In the case of $q_0$, the best fit value  for model D4,  $q_0=-0.637$,  is in agreement (at $1\sigma$ confidence level) with the $q_0$ value obtained from the MI approach. In this case, also the  $\Lambda$CDM result,  $q_0=-0.573$, and the D3 one,   $q_0=-0.549$,  are in agreement with the cosmographical results, but within $1-2\sigma$ uncertainty.  So both RVMs are fairly consistent with $q_0$. As for the GDEs, we find that while the D1 result, $q_0=-0.526$, is roughly in agreement at a similar level to the previous ones,  model  D2 exhibits the maximum tension with the MI results. Comparing the values of the jerk parameter, we observe that both $\Lambda$CDM with $j_0=1.0$ and D3 with $j_0=0.918$ are consistent with the MI ones at just $1\sigma$ confidence level.  In contrast, the computed values of  $j_0$ for three of the models (D1, D2 and D4) are far from the MI analysis, if using only this data source.  Thus, with the Pantheon sample,  models D1, D3 and D4 are consistent with  the $q_0$ value from the MI method, while only D3 gives a fairly consistent result with $j_0$.  From the results of this part, we conclude that $\Lambda$CDM can be falsified only at a moderate level of  $\sim 1.3 \sigma$  when using the  $q_0$ value. However, on using the jerk parameter, the $\Lambda$CDM model is still consistent with the MI result at $\sim 0.9\sigma$ level.  Obviously, no firm conclusion can be derived with only this data source.
These results are summarized in the upper-left panel of Fig.\ref{fig:miandmodels}. In this figure we have plotted the contours corresponding to the MI constraints in the $j_0-q_0$ plane up to $3\sigma$ confidence level, while the error bars display the computed value of the cosmographic parameters for the DDEs under study up to $1\sigma$ level.

\item {\bf Pantheon + BAOs:}  Adding another well-established sort of cosmological observations, namely the  BAOs data, to the previous SnIa sample we obtain  the Pantheon+BAOs  combined sample,  which can be used to derive the corresponding  MI  values of the cosmographic parameters.  We consider such a  combined set  as a most robust one in the sense that these data have been tested and should therefore provide the most reliable results before we use other data combinations.  The Pantheon+BAOs  robust sample renders  $q_0=-0.609\pm0.094$ and $j_0=0.892\pm0.44$ (cf. Table 1).    Let us  next confront these MI results  with the computed values of $q_0$ and $j_0$ ensuing from the best fit parameters  of the DDEs.  For the RVM models D4 (D3) we obtain $q_0=-0.619 (-0.606)$, which  are in full agreement with the MI values. The corresponding $q_0$ results for models D1 and D2 are only in moderate  $1 \sigma$ tension with the MI result. The concordance model value of $q_0$ deviates from the MI result by less than $0.5\sigma$.  So far so good. As for the  jerk parameter, we have an inconspicuous tension of about $1.1 \sigma$ for D3 and D4 with the MI case.
In the upper-right panel of Fig.\ref{fig:miandmodels}, the contour plots show the best fit values of the cosmographic parameters obtained using the Pantheon+BAOs sample and their confidence levels up to $1\sigma$. At the same time, we plot the best fit values of different DDEs for comparison (including the corresponding error bars). One can see that in the $q_0 - j_0$ plane, $\Lambda$CDM, D3 and D4 are in agreement with the MI result, whereas D1 and D2  are outside of the confidence regions. The net result from the  robust  Pantheon + BAOs sample is that the concordance $\Lambda$CDM model,  as well as the running vacuum models D3 and D4 are the most favored options, while the ghost dark energy models  D1 and  D2 appear already unfavorable in this crucial stage.

\item {\bf Pantheon + QSOs:}  Consider now the effect of adding QSOs data to the  Pantheon SnIa sample, without including BAOs data, just to check in a most clear way the effect of the quasars on the supernovae sample.  Using this combination, the MI approach renders $q_0=-0.844\pm0.048$ and $j_0=2.42\pm0.25$.   Both parameters have undergone a significant enhancement as compared to the previous case.  We may confront these results with the computed values of $q_0$ and $j_0$ as obtained from the different DE models according to their best fit parameters.  Insofar as $q_0$ is concerned,  D4 yields $q_0=-0.661$, which  is not very different from the previous theoretical value under Pantheon data alone, but now is in  $3.8\sigma$ tension with the MI value owing to the substantial increase of the latter caused by the QSOs. Despite the discrepancy, it is the less acute one among the different models. The other models deviate from the output of the MI analysis by at least $4.5\sigma$.  One could say that these results on the deceleration parameter are a bit chocking. The concordance $\Lambda$CDM, for example, turns out to be  in tension with $q_0$ at $\sim 6\sigma$.
In the case of the jerk parameter, we have a tension of more than $4.6 \sigma$ with the MI output.
In the bottom-left panel of Fig.\ref{fig:miandmodels}, the contours show the best fit values of the cosmographic parameters obtained using the combined Pantheon+QSO sample and their confidence levels up to $3\sigma$. At the same time, we plot the best fit values of different DE models for comparison with the corresponding error bars. We observe that in the $q_0 - j_0$ plane, all of the DDEs as well as standard $\Lambda$CDM are outside of the confidence regions. The net result from the   Pantheon + QSO combination is that D4 is the less disfavored model, while  D1, D2, D3 and $\Lambda$CDM are completely unfavorable. This is the first striking result, which might be due to the large error bars inherent to the  Hubble diagram data of QSOs  at higher redshifts. Notice that the Hubble diagram data of SNIa have smaller errors.  \jt{Notice that in \citep{Risaliti:2018reu}, the authors reported an internal discrepancy between QSOs data for redshifts less and greater than $1.4$.} \jt{In order to check the possible impact of this discrepancy in our case, we have repeated our analysis by considering a new (reduced) data sample which includes Pantheon plus QSOs only  in  the redshift range $z<1.4$.  Upon comparing the results of this subsample with those obtained using Pantheon alone  we find that the results for the MI case, as well as for $\Lambda$CDM, do not point to significant differences. For example,  within the MI approach using the Pantheon sample we obtain $q_0=-0.702\pm 0.104$ and $j_0=1.60 \pm 0.71$, while after adding only the low redshift QSOs, we obtain  $q_0=-0.699 \pm 0.108$ and $j_0=1.62 \pm 0.72$. In other words, the results  obtained from Pantheon+low redshift QSOs are completely consistent with the results of the Pantheon sample alone within $1\sigma$ confidence level.  In contrast,  in the case of the original Pantheon+full QSOs sample the results change significantly as compared to the Pantheon sample, as we had noted above. Thus, although our analysis confirms the mentioned internal discrepancy between the results obtained with quasars at  $z<1.4$ from  those obtained with quasars at $z> 1.4$, the effect of keeping only the low redshift QSOs with the Pantheon sample does not produce results different from using  the  Pantheon sample only.}

\item {\bf Pantheon + QSOs + GRBs:} A second unexpected result along similar lines  appears in the last step, when we combine the  three data sets. The corresponding MI results now read as follows:  $q_0=-0.819\pm0.065$ and $j_0=2.21_{-0.42}^{+0.37}$.  Comparing once more with the model-dependent computed values, we find that D3  yields $q_0=-0.608$  and it is now the less disfavored among the  DDEs,  although  still  $\sim 3.2 \sigma$ away from the MI determination.  The  other models yield results which  are in more than $4\sigma$ tension with respect to the MI ones. Therefore, in the case of Pantheon + QSOs + GRBs data combination, the results we obtain for $q_0$, indicate that none of the models are reasonably consistent with the MI approach. In the case of the  jerk parameter, as well as the deceleration parameter, model D3 has the best performance among different models. Here we obtain $j_0=1.289$, which is in moderate $2.2 \sigma$ tension with the MI result. In this case,  D1, D2 and D4 are the less favored.  Overall  model D3 is the best model in this case while we find tensions above $4\sigma$ for $q_0$ and also more than $3\sigma$ for $j_0$ for the other models. This means that assuming the high redshifts data points of GRBs, D1, D2 and D4 as well as $\Lambda$CDM are not supported.
In the bottom-right panel of Fig.\ref{fig:miandmodels}, we have shown the contour plots and confidence levels for  $q_0 - j_0$ obtained using the Pantheon+QSOs+GRBs data combination within the  MI approach, along with the computed values of the cosmographic parameters for the different DDEs.  As we can see,  the results on the deceleration and jerk parameters when BAOs data are replaced with QSOs or QSOs+GRBs data are  a bit perturbing, but as indicated above they might point to the fact that the QSOs and GRBs data are not sufficiently consolidated in comparison to other better-established cosmological probes, such as BAO data and Hubble parameter $H(z)$ measurements, and they may need some further qualification in order to explain the seeming inconsistencies found between the different results. As mentioned before, the puzzling results with the higher redshift data are nevertheless along the lines of previous (cosmographic and non-cosmographic) analyses involving the $\CC$CDM and other models\, \citep{Lusso:2019akb}  and \citep{Khadka:2019njj,Khadka:2021dsz}. Since GRBs probe a largely unexplored region of $z$, it is worth acquiring more and better-quality gamma-ray burst data before we can  offer  a more definitive answer as to weather they can provide  tests of cosmological models at a level of quality comparable to more traditional and well-established sets of observations. In the meantime, like the QSOs results, one may possibly attribute the significant  deviation between the computed  values of the cosmographic parameters from the MI approach and those from the DE models studied here to the large errors and uncertainties of GRBs at higher redhsifts.   Another possible interpretation of \textit{these deviations}  is that the Taylor expansions of the cosmographic analysis might not work optimally at the higher redshifts where QSOs, and especially the  GRBs, are observed.  For example,  the authors of \citep{Banerjee:2020bjq} claim that the cosmographic method based on the logarithmic expansion of the luminosity distance is only reliable up to redshifts $z\simeq 2$ (see also \cite{Yang:2019vgl}).  Hence,
if we would adopt this point of view,  any tension between the results of the cosmographic method stemming from the use of high redshift sources and the cosmological models might not be literally real.  On the other hand the authors of \cite{Bargiacchi:2021fow}, by using orthogonal logarithmic polynomials for the expansion of the luminosity distance, suggest that the cosmographic method provides a very good fit up to redshift  $z\simeq 7$. In such case the deviations between the cosmographic method and the cosmological models under study might be interpreted as  a real tension. However, in our case we have not used orthogonal logarithmic polynomials. Therefore, even though we mention here these considerations existing in the literature, we refrain from  adopting  a particular position at this point, except rendering the actual results of our analysis, of course. \jt{Furthermore, in the next section, in order to calculate the effect of error truncation of Taylor expansion at higher redshifts, we repeat our analysis using the mock data for Hubble diagrams of QSOs generated based on the standard $\Lambda$CDM cosmology.}
\end{enumerate}

Finally,  it is instructive to compare the computed value of $q_0$  from the DDEs under study with the value of the deceleration parameter which was found in \citep{Gomez-Valent:2018gvm} in a completely different MI approach. The value of the $q_0$ obtained from the reconstruction of $q(z)$ using the data on SnIa, CCH  and BAOs in the mentioned study, using the so-called Weighted Function Regression (WFR) method (previously developed in \cite{Gomez-Valent:2018hwc}), reads $q_0=-0.60 \pm 0.10$. It is interesting to see that the computed value of $q_0$ in our case from  the $\Lambda$CDM and also using the RVMs (models  D3 and D4) is in very good agreement with the result of \citep{Gomez-Valent:2018gvm}. On the other hand, the computed value of $q_0$  using the GDEs, especially  D2, is not in good agreement with the MI result of \citep{Gomez-Valent:2018gvm}. It is one more hint that the GDEs are difficult to reconcile with the cosmographic data, and in general with the overall cosmological observations (cf.\,\cite{Rezaei:2019xwo}).

\begin{figure*}
	\centering
	\includegraphics[width=8.5cm]{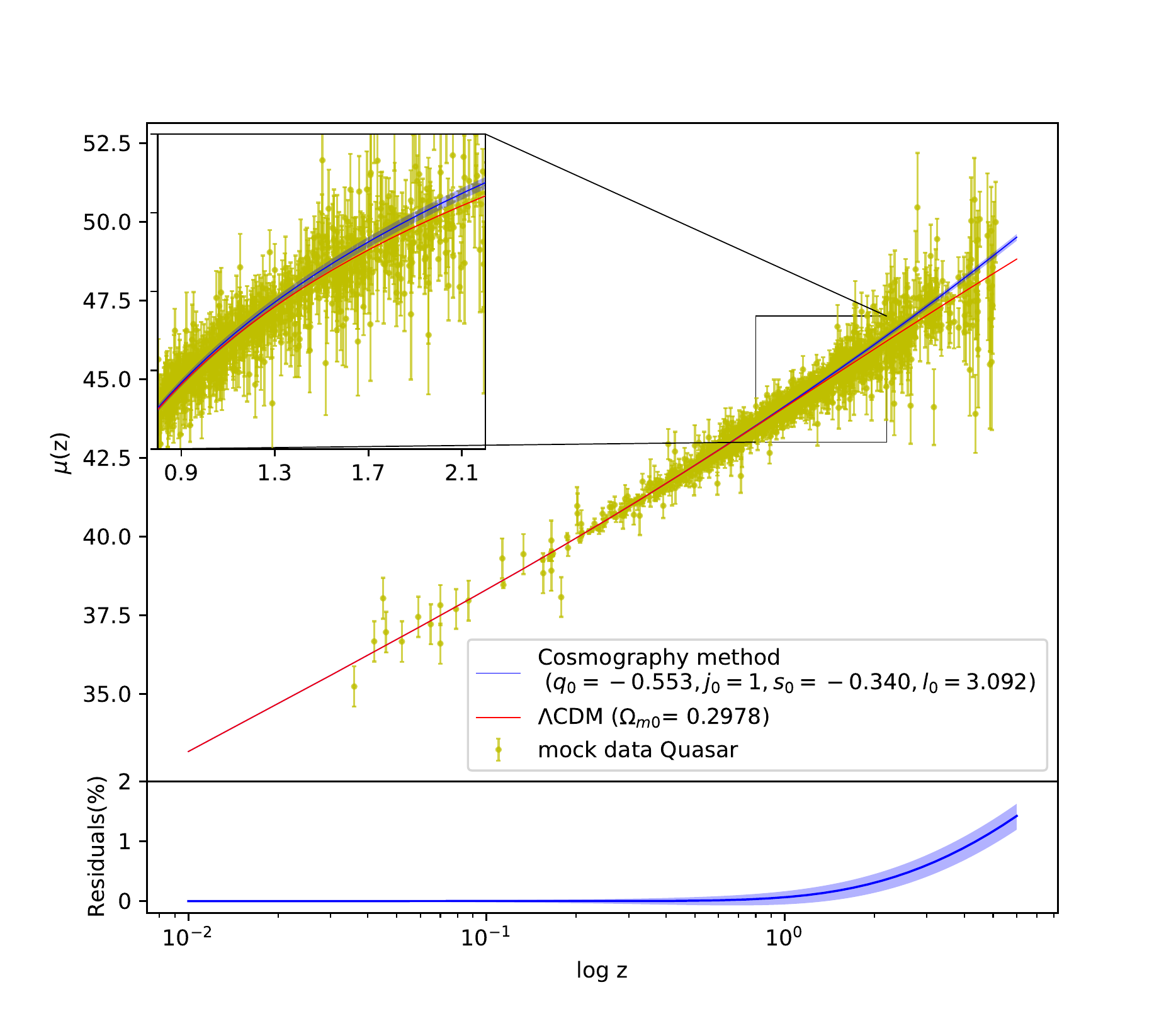}
	\caption{\jt{ Upper part: The Hubble diagram of QSOs in the context of standard flat $\Lambda$CDM model (red curve), the reconstructed Hubble diagram using the cosmographic parameters (blue band) and mock Hubble diagram data (yellow points with error bars) produced by flat $\Lambda$CDM model using canonical value $\Omega_{m0}=0.3$. Lower part: The percentage difference between the distance modulus obtained from flat $\Lambda$CDM model and cosmographic method in residual plane.}}
	\label{fig:mock}
\end{figure*}

\jt{ \section{Mock data and cosmographic method}\label{sec:truncation}
In this section we propose a method to estimate the truncation error in our cosmographic analysis.  In the context of flat $\Lambda$CDM model and by fixing $\Omega_{m0}=0.3$, we generate mock data for Hubble diagram of QSOs at redshifts in which the QSOs have been observed. Using the standard value of the distance modulus $\mu_{\Lambda}(z_i)$ at specific redshift $z_i$, we generate the mock Hubble diagram data $(\mu(z_i), \Delta\mu)$ where $\Delta \mu$ is the error bar of the data. The distance modulus $\mu(z_i)$ is chosen using the normal distribution around the fiducial
value $\mu_{\Lambda}(z_i)$ with $\Omega_{m0}=0.3$ and standard deviation given by the error $\Delta \mu (z_i)$. Notice that we set the
redshifts $z_i$ and error bars $\Delta \mu (z_i)$ to the observational redshifts and observational error bars of the real QSOs data. Having mock data in hand, we run the MCMC algorithm for the standard $\Lambda$CDM model and obtain the best fit value of the model parameter $\Omega_{m0}$ and then calculate the best fit values of the cosmographic parameters for the model. If the above procedure for generating mock data is acceptable, we expect that the best fit value of $\Omega_{m0}$ will be very close to the canonical value $0.3$. On the other hand, using the cosmographic parameters obtained in our analysis, we reconstruct the distance modulus of $\Lambda$CDM model in the context of the cosmographic method. In this formalism, any difference between reconstructed $\mu(z_i)$ from the cosmographic method and $\mu(z_i)$ as obtained from the exact Hubble parameter (with the best fit value of $\Omega_{m0}$) can be interpreted as a  truncation error of the Taylor series expansion used in the cosmographic approach. The results are shown in Fig.\,\ref{fig:mock}.  \jt{We quote also the best fit value of the mass density parameter,  $\Omega_{m}=0.2978^{+0.0094}_{-0.0094}$,  which agrees very well with its canonical value within $1\sigma$ confidence level}. The best fit values of the cosmographic parameters that we use to reconstruct the distance modulus in the context of cosmographic method are also indicated:  $q_0=-0.553\pm 0.014$, $j_0 = 1.00$, $s_0=-0.340\pm 0.042$ and $l_0=3.092^{+0.097}_{-0.11}$.
From Fig.\,\ref{fig:mock}, we observe that the deviation of the cosmographic method from the exact model starts to be appreciable  at $z\simeq 1.5$, meaning that \jt{at higher redshifts we encounter the first signs of the error truncation}. As we show in the residual plane of that figure, the percentage difference between the cosmographic approximation and the exact model is mild, specifically it is less than $1\%$ up to redshift $z\simeq 5$ and roughly $1.5 \%$ \jt{for the uppermost redshift range used in our analysis},  $5<z<6.7$.}

\section{Model selection criteria}\label{sec:selection}

In addition to testing models by considering their predictions on the cosmographical parameters, we use three other methods of model selection.   They take into account not only which model best minimizes the minimum value of  the  $\chi^2$ statistics but also the number of additional parameters of each model, and the corresponding penalty associated to the excess of parameters.  For example, in the standard $\CC$CDM model the number of parameters entering the present cosmographical analysis is $3$:  $(\Omega_{dm,0},  \Omega_{b,0}, h)$, namely the current dark matter and baryonic  densities (normalized with respect to the critical density)  and the reduced Hubble parameter:  $h=H_0/(100\,  {\rm  km}/s/{\rm Mpc})$.  As previously indicated, the DDEs under scrutiny add up $0$ (D1) or $1$ (D2,D3,D4)  parameters  to the three previous ones, where we took into account the setting $\mu=-\nu$  made in the case of model D4 so as to break degeneracies.  Obviously the presence of new parameters might improve the quality of the fit and hence Occam's razor suggests to duly penalize the models carrying more parameters.  This is  achieved through standard model  selection methods  called the information criteria.  Within this methodology we have the Akaike information criterion (AIC) \citep{Akaike:1974}, the Bayesian information criterion (BIC) \citep{Schwarz:1974} and the deviance information criterion (DIC)\citep{Spiegelhalter:2002yvw}, see \cite{Liddle:2007fy} for a review.
In particular, AIC and BIC are defined by the following new statistics:
\begin{eqnarray}
{\rm AIC} = \chi^2_{\rm min}+2k\;,\nonumber\\
{\rm BIC} = \chi^2_{\rm min}+k\ln N\,,
\end{eqnarray}
where $\chi^2_{\rm min}$ is the minimum value of $\chi^2_{\rm tot}$, $k$ is the number of free parameters in the model and $N$ is the total number of observational data points. Usually, $N\gg k$ and this has been assumed  to write the above form for the  AIC formula (in fact, it is actually the case in our analysis).  Let us also mention that the BIC criterion defined here can be made more sophisticated by using the Bayes factor, namely the ratio of marginal likelihoods (i.e. of  evidences) of the models under comparison. It is certainly a more rigorous statistics (but also a much more cumbersome quantity to handle from the computational point of view), see e.g. \cite{Sola:2018sjf} for recent applications of this method to cosmological model selection. However, an intermediate criterion which is easier to cope with and still benefits from the  direct use of the Markov chains of the MCMC analysis, is the DIC.  Such efficient tool employs both Bayesian statistics and information theory concepts and it is expressed as follows \citep{Spiegelhalter:2002yvw,Liddle:2007fy}:
\begin{eqnarray}
{\rm DIC} =\chi^2_{\rm min}+2p_D= 2\overline{\chi^2_{\rm tot}}-\chi^2_{\rm min}\;.
\end{eqnarray}
where $p_D=\overline{\chi^2_{\rm tot}}- \chi^2_{\rm min}$ is the so-called effective number of parameters of the model (as it is obvious from comparison with the AIC, of which the DIC is a more powerful generalization).  The value of $2p_D$ is sometimes called the model complexity.  The over-lines denote an average over the posterior distribution.  In a summarized way, DIC measures the sum of the deviation plus complexity of a given model. Using the results of our MCMC analysis for the different models under study, we evaluate these information criteria for each one of them. We present the  corresponding numerical results   for the different data combinations in Tables \ref{infop}-\ref{infopqg}. In order to sort out the models based on the value of information criteria, we have computed $\Delta {\rm AIC}$, $\Delta {\rm BIC}$ and $\Delta {\rm DIC}$ , in which $\Delta$ means the differences between the value of information criteria for each model, minus the corresponding value for the best model. According to the usual jargon,  small differences of $\Delta {\rm AIC}<2$ with respect to the best model, indicate still significant support to a given model versus the best model. Furthermore, we use $\Delta {\rm BIC}$ to gauge the evidence against a given model as compared to the best one \citep[for more details we refer the readers to][]{Kass:1995loi,Burnham2002,Rezaei:2019xwo}.  In  \citep{Burnham2002} it is suggested that models receiving a surplus of AIC within $1-2$ of the best model still deserve consideration, but if it is $3-7$ they have considerably less support. These rules of thumb appear to work reasonably well for DIC as well \citep{Spiegelhalter:2002yvw}.
\begin{table*}
 \centering
 \caption{The different information criteria obtained for  the DDE models and the concordance $\Lambda$CDM using Pantheon sample.
}
\begin{tabular}{c  c  c c c  c c c c c}
\hline \hline
Model & $\chi^2_{\rm min}$ & $k$ & $N$ & ${\rm AIC}$ & ${\rm BIC}$& ${\rm DIC}$ & $\Delta {\rm AIC}$ & $\Delta {\rm BIC}$ & $\Delta {\rm DIC}$\\
\hline
D1 & $1037.7$ & $3$ & $1048$ &  $1043.7$ & $1058.6$ &  $1043.3$  & $3.3$ & $2.0$ & $8.7$ \\
\hline
D2 & $1037.8$ & $4$ & $1048$ &  $1045.8$ & $1065.6$ &  $1043.6$  & $5.4$ & $9.1$ & $9.0$ \\
\hline
D3 & $1035.7$ & $4$ & $1048$ &  $1043.7$ & $1063.5$ &  $1040.1$  & $3.3$ & $7.0$ & $5.5$\\
\hline
D4 & $\textbf{1032.4}$ & $4$ & $1048$ &  $\textbf{1040.4}$ & $1060.2$ &  $\textbf{1034.6}$  & $\textbf{0.0}$ & $3.7$ & $\textbf{0.0}$ \\
\hline
$\Lambda$CDM & $1035.7$ & $3$ & $1048$ &  $1041.7$ & $\textbf{1056.6}$ &  $1037.8$  & $1.3$ & $\textbf{0.0}$ & $3.1$ \\
\hline \hline
\end{tabular}\label{infop}
\end{table*}

\begin{table*}
 \centering
 \caption{The different information criteria obtained for the DDE models and the concordance $\Lambda$CDM using Pantheon+BAOs sample.
}
\begin{tabular}{c  c  c c c  c c c c c}
\hline \hline
Model & $\chi^2_{\rm min}$ & $k$ & $N$ & ${\rm AIC}$ & ${\rm BIC}$& ${\rm DIC}$ & $\Delta {\rm AIC}$ & $\Delta {\rm BIC}$ & $\Delta {\rm DIC}$\\
\hline
D1 & $1043.4$ & $3$ & $1052$ &  $1049.4$ & $1064.3$ &  $1047.0$  & $6.3$ & $3.7$ & $10.3$ \\
\hline
D2 & $1042.3$ & $4$ & $1052$ &  $1050.3$ & $1070.1$ &  $1045.9$  & $7.2$ & $9.5$ & $9.2$ \\
\hline
D3 & $1036.1$ & $4$ & $1052$ &  $1044.1$ & $1063.9$ &  $1037.9$  & $1.0$ & $3.3$ & $1.2$\\
\hline
D4 & $\textbf{1035.1}$ & $4$ & $1052$ &  $\textbf{1043.1}$ & $1062.9$ &  $\textbf{1036.7}$  & $\textbf{0.0}$ & $2.3$ & $\textbf{0.0}$ \\
\hline
$\Lambda$CDM & $1039.7$ & $3$ & $1052$ &  $1045.7$ & $\textbf{1060.6}$ &  $1042.1$  & $2.6$ & $\textbf{0.0}$ & $5.4$ \\
\hline \hline
\end{tabular}\label{infopb}
\end{table*}

\begin{table*}
 \centering
 \caption{The different information criteria obtained for the DDE models and the concordance $\Lambda$CDM using Pantheon+QSO samples.
}
\begin{tabular}{c  c  c c c  c c c c c}
\hline \hline
Model & $\chi^2_{\rm min}$ & $k$ & $N$ & ${\rm AIC}$ & ${\rm BIC}$& ${\rm DIC}$ & $\Delta {\rm AIC}$ & $\Delta {\rm BIC}$ & $\Delta {\rm DIC}$\\
\hline
D1 & $1099.9$ & $3$ & $1073$ &  $1105.9$ & $1120.8$ &  $1136.3$  & $6.5$ & $6.3$ & $40.0$ \\
\hline
D2 & $1099.5$ & $4$ & $1073$ &  $1107.5$ & $1127.4$ &  $1122.9$  & $8.1$ & $13.0$ & $27.0$ \\
\hline
D3 & $1096.3$ & $4$ & $1073$ &  $1104.3$ & $1124.2$ &  $1099.5$  & $4.9$ & $9.7$ & $3.7$\\
\hline
D4 & $\textbf{1092.4}$ & $4$ & $1073$ &  $\textbf{1099.4}$ & $1119.3$ &  $\textbf{1095.8}$  & $\textbf{0.0}$ & $5.8$ & $\textbf{0.0}$ \\
\hline
$\Lambda$CDM & $1093.6$ & $3$ & $1073$ &  $1099.6$ & $\textbf{1114.5}$ &  $1100.8$  & $0.2$ & $\textbf{0.0}$ & $5.0$ \\
\hline \hline
\end{tabular}\label{infopq}
\end{table*}

\begin{table*}
 \centering
 \caption{The different information criteria obtained for the DDE models and the concordance $\Lambda$CDM using Pantheon+QSOs+GRBs samples.
}
\begin{tabular}{c  c  c c c  c c c c c}
\hline \hline
Model & $\chi^2_{\rm min}$ & $k$ & $N$ & ${\rm AIC}$ & ${\rm BIC}$& ${\rm DIC}$ & $\Delta {\rm AIC}$ & $\Delta {\rm BIC}$ & $\Delta {\rm DIC}$\\
\hline
D1 & $1317.6$ & $3$ & $1073$ &  $1323.6$ & $1339.0$ &  $1324.2$  & $5.8$ & $3.6$ & $13.2$ \\
\hline
D2 & $1317.1$ & $4$ & $1235$ &  $1325.1$ & $1345.6$ &  $1322.9$  & $7.3$ & $10.0$ & $11.9$ \\
\hline
D3 & $1310.2$ & $4$ & $1235$ &  $1318.2$ & $1338.7$ &  $1311.6$  & $0.4$ & $3.3$ & $0.6$\\
\hline
D4 & $\textbf{1309.8}$ & $4$ & $1235$ &  $\textbf{1317.8}$ & $1338.3$ &  $\textbf{1311.0}$  & $\textbf{0.0}$ & $2.9$ & $\textbf{0.0}$ \\
\hline
$\Lambda$CDM & $1314.0$ & $3$ & $1235$ &  $1320.0$ & $\textbf{1335.4}$ &  $1315.9$  & $2.2$ & $\textbf{0.0}$ & $4.9$ \\
\hline \hline
\end{tabular}\label{infopqg}
\end{table*}
%%%%%%%%%%%%%%%%%%%%%%%%%%%%%%%%%%%%%%%%%%%%%%%%%%%%%%%%%%%%%%%%%%%%%%%%%%%%%%%%%%%%%
For each of the criteria, we  have displayed the results of the best model with bold fonts in Tabs.(\ref{infop}-\ref{infopqg}).
%%%%%%%%%%%%%%%%%%%%%%%%%%%%%%%%%%%%%%%%%%%%%%%%%%%%%%%%%%%%%%%%%%%%%%%%%%%%%%%%%%%%%
    \begin{table*}
  \centering
  \caption{The support to (or evidence against) each DE model obtained using the results from the different information criteria.
 }
 \begin{tabular}{c  c  c c c}
 \hline \hline
Data & Model &  {\rm AIC} & {\rm BIC} & {\rm DIC}\\
 \hline
  & D1 &  Considerably less support  &  Mild to positive evidence  & Essentially no support  \\
  & D2 &  Considerably less support  & Strong evidence & Essentially no support  \\
 Pantheon & D3 &  Considerably less support  & Strong evidence  & Considerably less support  \\
  & D4 & \bf{Best model}   &   Mild to positive evidence  & \bf{Best model}  \\
  & $\Lambda$CDM & \bf{Significant support}   &  \bf{Best model}  & Considerably less support  \\
 \hline
    & D1 &   Considerably less support  &  Mild to positive evidence  & Essentially no support  \\
     & D2 &   Essentially no support  & Strong evidence & Essentially no support  \\
    Pantheon+BAOs & D3 &  \bf{Significant support}  & No evidence  & \bf{Significant support}  \\
     & D4 & \bf{Best model}   &   Mild to positive evidence  & \bf{Best model}  \\
     & $\Lambda$CDM & Considerably less support   &  \bf{Best model}  & Considerably less support  \\
  \hline
    & D1 &  Considerably less support  &  Strong evidence  & Essentially no support  \\
    & D2 &  Essentially no support  &  Very strong evidence  & Essentially no support  \\
 Pantheon+QSOs& D3 & Considerably less support   & Strong evidence   &  Considerably less support \\
    & D4 & \bf{Best model}   &  Mild to positive evidence  &  \bf{Best model} \\
    & $\Lambda$CDM & \bf{Significant support}   &  \bf{Best model}  &  Considerably less support \\
   \hline
      & D1 &  Considerably less support  &  Mild to positive evidence   & Essentially no support  \\
      & D2 & Essentially no support   & Strong evidence   & Essentially no support  \\
 Pantheon+QSOs+GRBs & D3 &  \bf{Significant support}  &   Mild to positive evidence  & \bf{Significant support}  \\
      & D4 &  \bf{Best model}  &   Mild to positive evidence  &  \bf{Best model}  \\
      & $\Lambda$CDM & Considerably less support   &  \bf{Best model}   &  Considerably less support  \\
    \hline \hline
   \end{tabular}\label{chartinf}
   \end{table*}
In Table\,\ref{chartinf} we have summarized the verdict of the different information criteria for the DE models under consideration using each of the data combinations. In that convenient table one can read off  immediately  the support to (or  evidence against) each model obtained using different data combinations. \jt{We note that the DIC criteria is a generalization of AIC, in which we use the  Kullback-Leibler divergence instead of the squared error loss. Hence, one can expect that the results of AIC and DIC are approximately close. \jt{In fact, the complexity parameter $2p_D$  in DIC is the direct analog of  $2k$, twice the effective number of free parameters in AIC.  From Table 11 we can see that there is an excellent resonance in picking out the best model either using AIC or  DIC.}  On the other hand, \jt{the BIC criterion} provides a convenient approximation, which may be interpreted as a penalized maximum likelihood corresponding to a given model. This criteria is more sensitive to the free parameters and its results can be different from AIC and DIC criteria. See e.g. \cite{Liddle:2007fy} for more details.}

From Table\ref{chartinf},  we find that the $\Lambda$-cosmology  ($\CC$CDM) is reckoned from the {\rm BIC} point of view as being consistent with all of the data combinations at both low and high redshifts, but it is not the best option. Among the different DDEs,  the running vacuum models  D3 and D4 are the best positioned ones. Globally, the D4 model obtains  a pretty good mark from the cosmographic point of view as compared to the rest, both from the perspective of the AIC and the DIC,  whereas  both GDEs,  D1 and D2 get a bad score since we find strong evidence against them,  all the more when we take into account the high redshift data. It means that the GDEs are not supported by the cosmographical analysis.  Let us emphasize that these models are not supported by the overall cosmological analysis either (in which BAOs, LSS and CMB were included), see  \citep{Rezaei:2020lfy}. In stark contrast with the GDEs, in  this same reference we have shown that the RVMs are indeed supported by the LSS and CMB data,  together with  BAOs,  and in particular we have stressed that the RVMs are able to alleviate the $H_0$ and $\sigma_8$ tensions, see also the recent work \cite{Sola:2021txs}.  For this reason we may conclude that the RVMs are particularly favored from the optic of  both the local cosmographic and global cosmological analyses.

\section{Conclusions} \label{conlusion}

In this work, using different combinations of Hubble diagram data sets  for Pantheon (SnIa), QSOs (quasars) and GRBs (gamma rays bursts), as well as BAOs (baryonic acoustic oscillations), we have compared different dynamical DE models (DDEs) with the concordance $\Lambda$CDM model from the viewpoint of the cosmographic approach. For such purpose, we have utilized different methods for model comparison including a variety of information criteria. Two scenarios for dynamical DE have been investigated whose energy densities can be expressed as a power series expansion of the Hubble rate and its time derivatives, to wit: the Ghost DE models (GDEs) and the running vacuum models (RVMs). The data points used in our analysis  cover a wide range of redshifts, from $z=0.01$ to $z=6.67$, which enable us to investigate the evolution of the universe in the presence of DE. Assuming different combinations of data sets we have studied the phenomenological performance of the mentioned DE models in different redshift ranges.
The GDEs come out to be the less favored models  in our analysis. They appear to be so already in the first stage when we just use the Pantheon sample of supernovae.  Using this sample, and above all when upgrading it with BAOs, the  combined Pantheon+BAOs dataset can be fitted with the  $\Lambda$CDM and the RVMs and the results prove to be  in fairly good agreement with those from the cosmographic MI (model independent) approach.    Intriguingly enough,  however, with  the Pantheon+QSO  and  Pantheon+QSO+GRB datasets we find that essentially  all of the DE models look rather disfavorable, being model D3 the less tensioned one of them.  As mentioned in the main text, insofar as concerns the $\CC$CDM model, this striking situation is in full agreement with the results of \citep{Lusso:2019akb}, which confirm the tension between the best fit cosmographic parameters and the $\Lambda$-cosmology at a significant level, chiefly with the full SnIa+QSOs+GRBs data set.  However, as pointed out in the main text,  one cannot exclude that a possible origin of such tension may be caused by the convergence problem of the Taylor series expansion  for the sources characterized by the highest redshifts used in our cosmographic samples. This possibility should be further  examined by using alternative methods of  MI analysis. From the cosmographic perspective,  the tensions between the best MI values of the deceleration and jerk parameters, $q_0$ and $j_0$,  and the computed values from $\Lambda$CDM loosen as soon as  the  Pantheon or  Pantheon+BAOs data  are employed.   The upshot is that from the point of view of  the cosmographic approach the standard $\CC$CDM model and the RVMs (models D3 and D4)  are consistent with the  Pantheon+BAO observational points, which can be considered the best-established cosmological probes. This is no longer true, however,  when  BAOs are replaced with QSOs and/or GRBs. In such case  all the models have serious difficulties to fit the MI data, but above all the GDEs.  We cannot exclude that these troubles may be  due to the larger observational errors afflicting the data at the higher redshifts explored in our analysis.

Fortunately, from the standpoint of the  information criteria we find a more definite picture.   The concordance $\Lambda$CDM model proves consistent with the  Pantheon, Pantheon+BAOs and Pantheon+QSO data combinations. Especially when using the {\rm BIC} criterion,  the $\Lambda$CDM appears as the best model for all the data combinations.  But BIC is only a rough estimate of the Bayes factor, i.e. the ratio of marginal likelihoods. If, instead,  we use the deviance information criterion ({\rm DIC}),  which is perhaps the most sophisticated of our information criteria since it  makes direct use of the Markov chains entering our statistical analysis, the most favored model turns out to be  D4  by far, followed by D3 in most cases.  In other words,  the two RVMs are the most favored models by the DIC criterion.   This conclusion is virtually in accordance with the verdict of the basic {\rm AIC} criterion as well. As for the two ghost dark energy models, D1 and D2,  they become consistently ruled out also by (all) the information criteria  used in this study.  Let us remark that this is in full agreement with the results which we obtained in \citep{Rezaei:2019xwo} using cosmological data sets beyond the cosmographic approach (e.g.  involving  structure formation data and CMB).  In all of the data combinations, we find that the running vacuum model D4 (closely followed by D3)  provides the best global fit  as compared to the other DDEs.  Finally, we should recall from the results of the aforementioned reference that models D3 and D4 have both the capacity to  alleviate the two famous $\sigma_8$ and $H_0$ tensions mentioned in the introduction.

Here we have tested the running vacuum models  (D3 and D4) using only the cosmographic methodology and hence utilizing a limited portion of the data at low redshift, namely data which can be fully treated within the cosmographical method. Even with only these data we have confirmed that  the RVMs  perform better than the $\CC$CDM and also far better than other DDEs which we have used for the sake of comparison (the GDEs).  Let us also note that despite the fact that simple parameterizations of the DDE,  such as the well-known wCDM (also called  XCDM) \citep{Turner:1998ex}  or  CPL\,\citep{Chevallier2001,Linder:2002et}, can also perform well at the level of cosmography as compared to the $\CC$CDM (cf. the analysis of \cite{Rezaei:2020lfy}), it turns out that these  parameterizations (which are ultimately phenomenological by nature) are unable to provide a minimal explanation for the severe $H_0$ tension, as clearly shown  e.g. in \cite{Sola:2018sjf}.

Summarizing, we have demonstrated  that the class of running vacuum models (RVMs) possess several interesting phenomenological capabilities at different levels, which definitely help to improve the description of the cosmographic and in general the cosmological observations.  This is remarkable if we take into account that they are rooted in a sound theoretical  QFT/string framework (cf. \citep{Sola:2007sv,Sola:2013gha,Sola:2015rra,Mavromatos:2020kzj}. On the  phenomenological side,  they can pass the cosmographic tests,  improve the overall fit to different kinds of cosmological data at both low and high redshift,  and at the same time they can help to mitigate the persistent conflict with the $\sigma_8$ and $H_0$ tensions\,\citep{Rezaei:2020lfy,Sola:2021txs}. Finally, we note that the puzzling inconsistencies between the  cosmographic approach and the information criteria might be attributed to the present status of the QSO+GRB data, which may require some additional subsampling and/or refinement before they can be put on equal footing with  the robust SnIa+BAO data.  This is, however,  beyond the scope of the present study.

\section{Acknowledgements}
One of the authors, JSP, acknowledges partial support by  projects  PID2019-105614GB-C21 and FPA2016-76005-C2-1-P (MINECO, Spain), 2017-SGR-929 (Generalitat de Catalunya) and CEX2019-000918-M (ICCUB). JSP also acknowledges participation in the COST Association Action CA18108 ``{\it Quantum Gravity Phenomenology in the Multimessenger Approach (QG-MM)}''.

\section{Data availability}

The data underlying this article will be shared on reasonable request to the corresponding author.

 \bibliographystyle{mnras}

\end{document}